%% file: peaks.tex
\documentclass[a4paper,11pt]{article}
\pdfoutput=1

\usepackage{jcappub}

\usepackage[T1]{fontenc}

\usepackage{aas_macros}
\usepackage{slashed}

\title{\boldmath The shape of CMB temperature and polarization peaks on the sphere}

\author[a,b]{A.~Marcos-Caballero,}
\author[a]{R.~Fern\'andez-Cobos,}
\author[a]{E.~Mart\'\i nez-Gonz\'alez}
\author[a]{and P.~Vielva}

\affiliation[a]{Instituto de F\'isica de Cantabria, CSIC-Universidad de Cantabria,\\ Avda. de los Castros s/n, 39005 Santander, Spain.}
\affiliation[b]{Dpto. de F\'isica Moderna, Universidad de Cantabria,\\ Avda. los Castros s/n, 39005 Santander, Spain.}

\emailAdd{marcos@ifca.unican.es}
\emailAdd{cobos@ifca.unican.es}
\emailAdd{martinez@ifca.unican.es}
\emailAdd{vielva@ifca.unican.es}

\abstract{\input{abstract}}

\begin{document}
\maketitle
\flushbottom

\input{body}

\acknowledgments
Partial financial support from the Spanish Ministerio de Econom\'{i}a
y Competitividad Projects AYA2012-39475-C02-01 and Consolider-Ingenio
2010 CSD2010-00064 is acknowledged.

\bibliographystyle{JHEP}
\bibliography{peaks.bib}

\end{document}

%% file: abstract.tex
We present a theoretical study of CMB temperature peaks, including its
effect over the polarization field, and allowing nonzero
eccentricity. The formalism is developed in harmonic space and using
the covariant derivative on the sphere, which guarantees that the
expressions obtained are completely valid at large scales (i.e., no
flat approximation). The expected patterns induced by the peak, either
in temperature or polarization, are calculated, as well as their
covariances. It is found that the eccentricity introduces a
quadrupolar dependence in the peak shape, which is proportional to a
complex bias parameter $b_\epsilon$, characterizing the peak asymmetry
and orientation. In addition, the one-point statistics of the
variables defining the peak on the sphere is reviewed, finding some
differences with respect to the flat case for large peaks. Finally, we
present a mechanism to simulate constrained CMB maps with a particular
peak on the field, which is an interesting tool for analysing the
statistical properties of the peaks present in the data.

%% file: body.tex
\section{Introduction}

The cosmic microwave background (CMB) radiation is one of the most
important sources of cosmological information. In particular, the
statistical properties of the CMB fluctuations are essential to
understand the primordial Universe. In order to explain the
observations, a phase of inflationary expansion in the early Universe
has been postulated. Within the standard frame, this inflation
mechanism also generates the initial matter perturbations which are
the seeds of the cosmic structures observed nowadays. It is believed
that the initial perturbations generated by the standard inflationary
models are nearly Gaussian. For this reason, the 2-point correlation
functions of the temperature and polarization CMB anisotropies have
most of the cosmological information which can be obtained from the
primordial fluctuations. The temperature angular power spectrum of the
CMB has been recently determined by the Planck mission with high
accuracy \cite{planck112015}. Regarding the primordial CMB
polarization, only the gradient part of the polarization field
($E$-mode) has been detected up to $\ell \sim 2000$. Although the
effect of the gravitational lensing on the curl of the polarization
field ($B$-mode) has been observed \cite{SPT2013,POLARBEAR2014}, there
is still no evidence of the primordial $B$-mode induced by the tensor
perturbations \cite{bicep-planck2015,bicep2015}). On the whole, the
agreement of the CMB data with the cosmological standard model is high
\cite{planck132015}. However, there are several anomalies at large
scales which are still unexplained \cite{planck162015}.

One of the alternative observables which can be used to study the
perturbations is the statistical properties of peaks. In the case of
the matter field, it is important to understand the properties of
overdensity peaks because that is where the collapse of structures
takes place. In a seminal work \cite{BBKS1986}, the statistics
of peaks for Gaussian fields in three dimensions is developed. There
are several aspects of peaks which can be analysed, for instance, the
number of peaks, the peak shape or their correlation function. One
important result from peak theory is that the peak correlation
function is related to the underlying matter distribution through a
non-local bias \cite{desjacques2013}. The understanding of the peak
correlation function and its bias relation to the matter field is also
important to study the baryon acoustic oscillations
\cite{desjacques2008}.

In order to study the CMB temperature extrema, the three-dimensional
formalism of peaks was later particularised in \cite{bond1987} to the
case of scalar fields on the sphere (see also
\cite{barreiro1997}). However, a full analysis including polarization
is needed for a complete understanding of the CMB fluctuations. The
radial profiles of the Stokes parameters were described in
\cite{komatsu2011}. Nevertheless, these profiles are calculated using
the small-angle limit and the peaks are considered spherically
symmetric. Recently, an analysis of the CMB temperature and
polarization Planck data, including peak eccentricity, has been
published \cite{planck162015}. The studies in \cite{komatsu2011} and
\cite{planck162015} based on the stacking of peaks do not reveal
significant deviations from the standard model, except a shift in the
temperature profile which could be associated to the power deficit at
large scales. Non-standard scenarios including parity violations
\cite{contaldi2015} or cosmological birefringence \cite{galaverni2015}
can also be tested using the stacking of temperature peaks in
polarization.

In this paper, we present a comprehensive study of the CMB peaks on
the sphere including polarization and allowing different
eccentricities. The derivation followed in this work is based on the
spherical harmonic coefficients, instead of the real space. This
allows to obtain expressions which are completely valid at large
scales, where the flat approximation breaks. In addition, the
formalism in harmonic space opens the possibility of generating
constrained CMB simulations with a peak at some point of the sphere
with the desired characteristics. Besides the peak shapes, the extrema
statistics is reviewed for the case of a Gaussian scalar field on the
sphere. It is found that the probability and the number density of
large peaks is modified with respect to the calculations in
\cite{bond1987}. Finally, we notice that the approach addressed in
this work is completely general and it can be applied to any scalar
Gaussian field on the sphere, taking into account its correlation with
any other scalar or spin-$2$ field.

This paper is organized as follows: in section~\ref{sec:derivatives},
we introduce the covariant derivatives in terms of the spherical
harmonic coefficients in order to define the peak degrees of freedom
on the sphere. In section~\ref{sec:uncorrelating}, it is explained the
methodology used to separate the variables defining the peak from the
rest of the Gaussian random field. Additionally, the statistics of
extrema on the sphere is reviewed in section~\ref{sec:extrema_stat},
finding some differences with respect to previous calculations. The
shape of CMB peaks including polarization for different values of mean
curvature and eccentricity is analysed in
section~\ref{sec:multipolar}, whilst its covariance is calculated in
section~\ref{sec:covariance}. The expressions of the peak patterns are
given in terms of the angular power spectra, which allows to calculate
them in a simple way. The physical description of the peak profiles is
discussed in section~\ref{sec:physical}. Furthermore, in
section~\ref{sec:simulations}, a way to simulate peaks on the sphere,
which is one of the applications of the formalism developed in this
work, is derived. Finally, the conclusions of the paper are presented
in section~\ref{sec:conclusions}.

\section{Derivatives of a scalar field on the sphere}
\label{sec:derivatives}

A peak on the sphere is defined through its derivatives up to second
order. In general, any field on the sphere can be expanded in terms of the
spherical harmonics:
\begin{equation}
T(\theta,\phi) = \sum_{\ell=0}^\infty a_{\ell m} \ Y_{\ell
  m}(\theta,\phi) \ .
\label{eqn:spherical_harmonic_expansion}
\end{equation}
The first step in our analysis is to express the derivatives in terms
of the spherical harmonic coefficients $a_{\ell m}$. For simplicity,
we consider that the peak is located at the north pole. The value of
the field at this point can be written in the following way:
\begin{subequations}
\label{eqn:peak_der}
\begin{equation}
T = \sum_{\ell=0}^\infty \sqrt{\frac{2\ell+1}{4\pi}} a_{\ell 0} \ .
\label{eqn:temp}
\end{equation}
In order to calculate derivatives on the sphere, we use the spin
raising and lowering operators $\slashed{\partial}$ and
$\slashed{\partial}^*$, which are proportional to the covariant
derivatives in the helicity basis (see appendix~\ref{app:cov_der} for
a more detailed description of the derivatives on the sphere). If we
consider the local system of reference at any point of the sphere,
then the derivatives with respect to the Cartesian coordinates
correspond to the real and imaginary parts of the lowering operator,
that is $\slashed{\partial}^* = -\partial_x + i \partial_y$
(similarly, the spin raising operator verifies $\slashed{\partial} =
-\partial_x - i \partial_y$). Here, we have assumed that the basis
vectors $\mathbf{e}_x$ and $\mathbf{e}_y$ correspond to the vectors
$\mathbf{e}_\theta$ and $\mathbf{e}_\phi$ of the spherical coordinate
system, respectively. Thus, the first derivatives at this point are
written as
\begin{equation}
\slashed{\partial}^* T = \sum_{\ell=0}^\infty
\sqrt{\frac{2\ell+1}{4\pi}} \sqrt{\frac{(\ell+1)!}{(\ell-1)!}} a_{\ell
  1} \ .
\label{eqn:temp_der}
\end{equation}
This quantity is a complex number whose real and imaginary parts
correspond to the derivatives in each orthogonal direction at the
north pole. Finally, the second derivatives are encoded in the Hessian
matrix (see eq.~\eqref{eqn:hessian}). It is convenient to separate the
trace and the traceless parts of this matrix, because these two
quantities transform in a different way under rotations. The trace
corresponds to the Laplacian,
\begin{equation}
\nabla^2 T = \slashed{\partial}^* \slashed{\partial} T = - \sum_{\ell=0}^\infty
\sqrt{\frac{2\ell+1}{4\pi}} \frac{(\ell+1)!}{(\ell-1)!} a_{\ell 0} \ ,
\end{equation}
and the traceless part is given by
\begin{equation}
(\slashed{\partial}^{*})^2 T = \sum_{\ell=0}^\infty \sqrt{\frac{2\ell+1}{4\pi}}
  \sqrt{\frac{(\ell+2)!}{(\ell-2)!}} a_{\ell 2} \ .
\label{eqn:D2psi}
\end{equation}
\end{subequations}
In the local system of reference, this operator is given by
$(\slashed{\partial}^{*})^2 = \partial_x^2-\partial_y^2 -
i2\partial_x\partial_y$. Although this operator is a complex quantity,
the imaginary part can be set to zero with a rotation of the $xy$
plane. Physically, this corresponds to choose the principal axes of
the peak as the reference system. From the real part, it can be shown
that the operator $(\slashed{\partial}^{*})^2$ represents a measure of
the anisotropy at the centre of the peak.

The $a_{\ell m}$ coefficients are $m$-spin quantities under rotations
of the $z$ axis. That is, if we rotate by an angle $\alpha$, the
$a_{\ell m}$ coefficient transforms as $a_{\ell m}
e^{im\alpha}$. Looking at the expression of the field derivatives in
terms of the spherical harmonic coefficients, it is possible to deduce
that, whilst $T$ and $\nabla^2T$ are scalars, $\slashed{\partial}^* T$
is a vector and $(\slashed{\partial}^*)^2 T$ is a 2-spin tensor. Since
tensors with different rank are statistically independent under the
assumption of isotropy, then only the scalars $T$ and $\nabla^2 T$
are correlated, while $\slashed{\partial}^*T$ and
$(\slashed{\partial}^*)^2T$ are uncorrelated with the rest of the
field derivatives.

For simplicity, we normalize the field derivatives in order to have
unit variance:
\begin{subequations}
\begin{equation}
\nu \equiv \frac{T}{\sigma_\nu} \ ,
\qquad
\kappa \equiv - \frac{\nabla^2T}{\sigma_\kappa} \ ,
\end{equation}
\begin{equation}
\eta \equiv \frac{\slashed{\partial}^*T}{\sigma_\eta} \ ,
\qquad
\epsilon \equiv \frac{(\slashed{\partial}^*)^2T}{\sigma_\epsilon} \ ,
\end{equation}
\end{subequations}
where the expressions of the variances are given in the
appendix~\ref{app:peak_dof}. These parameters denote the peak degrees
of freedom throughout the paper. The parameter $\nu$ represents the
peak height, whereas the normalized Laplacian $\kappa$ is the mean
curvature of the peak. The parameter $\eta$ is a complex number whose
components are the first derivatives at the peak location. Hereafter,
we set $\eta = 0$ in order to have a critical point. Finally, the
value of $\epsilon$ gives information about the eccentricity of the
peak. In particular, its modulus is proportional to the square of the
eccentricity, and its phase is twice the orientation angle with
respect to the reference system (see more details in
appendix~\ref{app:peak_dof}).

Using eqs.~\eqref{eqn:peak_der}, the peak variables can
be expanded in terms of the spherical harmonic coefficients, which are
normalized to have unit variance:
\begin{subequations}
\begin{equation}
\nu = \sum_{\ell=0}^{\infty} \nu_{\ell} a_{\ell 0} \ ,
\qquad
\kappa = \sum_{\ell=0}^{\infty} \kappa_{\ell} a_{\ell 0} \ ,
\label{eqn:nu_kappa}
\end{equation}
\begin{equation}
\eta = \sum_{\ell=0}^{\infty} \eta_{\ell} a_{\ell 1} \ ,
\qquad
\epsilon = \sum_{\ell=0}^{\infty} \epsilon_{\ell} a_{\ell 2} \ ,
\label{eqn:eta_epsilon}
\end{equation}
\end{subequations}
where the multipolar coefficients $\nu_\ell$, $\kappa_\ell$,
$\eta_\ell$ and $\epsilon_\ell$ are defined in the
appendix~\ref{app:peak_dof} (eqs.~\eqref{eqn:peak_dof}).

\section{Uncorrelating the peak variables}
\label{sec:uncorrelating}

The aim of this section is to separate the peak degrees of freedom
from the rest of the information of the field. For this purpose, we
transform the $a_{\ell m}$ coefficients into a new set of variables
containing the peak degrees of freedom ($\nu$, $\kappa$, $\eta$ and
$\epsilon$) and an ensemble of new variables $\hat{a}_{\ell m}$
without any peak information. The $\hat{a}_{\ell m}$ variables are
defined for all values of $\ell$ except for four given multipoles
$\ell_{\nu}$, $\ell_{\kappa}$, $\ell_{\eta}$ and $\ell_{\epsilon}$, in
order to preserve the total number of degrees of freedom.\footnote{In
  principle, the multipoles $\ell_{\nu}$, $\ell_{\kappa}$,
  $\ell_{\eta}$ and $\ell_{\epsilon}$ are chosen arbitrarily with the
  condition that $\nu_{\ell_\nu}, \kappa_{\ell_\kappa},
  \eta_{\ell_\eta}, \epsilon_{\ell_\epsilon} \neq 0$, such that the
  change of variables is not singular.} We choose the variables
$\hat{a}_{\ell m}$ such that its correlation with the peak variables
vanishes, using an orthogonalization process. For convenience, we
normalize the $a_{\ell m}$ coefficients such that they have unit
variance. The change of variables is given by:
\begin{subequations}
\label{eqn:alms}
\begin{equation}
\hat{a}_{\ell 0} = a_{\ell 0} -
\left(\begin{array}{cc} a_{\ell_{\nu} 0} & a_{\ell_{\kappa} 0} 
\end{array} \right)
P^{-1}
\left(\begin{array}{c} \nu_{\ell} \\ \kappa_{\ell} 
\end{array} \right) \quad (\ell \neq \ell_{\nu}, \ell_{\kappa}) \ ,
\label{eqn:al0}
\end{equation}
\begin{equation}
\hat{a}_{\ell 1} = a_{\ell 1} - a_{\ell_{\eta} 1}
\frac{\eta_{\ell}}{\eta_{\ell_{\eta}}} \quad (\ell \neq \ell_{\eta}) \ ,
\end{equation}
\begin{equation}
\hat{a}_{\ell 2} = a_{\ell 2} - a_{\ell_{\epsilon} 2}
\frac{\epsilon_{\ell}}{\epsilon_{\ell_{\epsilon}}} \quad (\ell \neq \ell_{\epsilon})
\ ,
\end{equation}
\begin{equation}
\hat{a}_{\ell m} = a_{\ell m} \quad (m > 2) \ .
\label{eqn:alm}
\end{equation}
\end{subequations}
The peak variables only affect to the multipoles $m = 0,1,2$, and
therefore the $a_{\ell m}$ coefficients with $m > 2$ remain
unchanged. The matrix $P$ in eq.~\eqref{eqn:al0} is the pivot matrix
given by
\begin{equation}
P = \left( \begin{array}{cc}
\nu_{\ell_{\nu}} & \nu_{\ell_{\kappa}} \\
\kappa_{\ell_{\nu}} & \kappa_{\ell_{\kappa}}
\end{array} \right) \ .
\label{eqn:pivot_matrix}
\end{equation}
Notice that the variables $\hat{a}_{\ell m}$ are not the coefficients
of the standard spherical harmonics expansion. The inverse relations
between $a_{\ell m}$ and $\hat{a}_{\ell m}$ are calculated from
eqs.~\eqref{eqn:alms} with a little bit of algebra:
\begin{subequations}
\label{eqn:alms_inv}
\begin{equation}
a_{\ell 0} =
\hat{a}_{\ell 0} + \left( \begin{array}{cc}
  \nu_{\ell} & \kappa_{\ell}
\end{array} \right) \Sigma^{-1} \left[
\left( \begin{array}{c}
\nu \\ \kappa
\end{array} \right) - 
\sum_{\ell^\prime=0}^\infty
\left( \begin{array}{c}
\nu_{\ell^\prime} \\ \kappa_{\ell^\prime}
\end{array} \right) \hat{a}_{\ell^\prime 0} \right] \ ,
\label{eqn:al0_inv}
\end{equation}
\begin{equation}
a_{\ell 1} = \hat{a}_{\ell 1} + \eta_{\ell} \left( \eta -
\sum_{\ell^\prime=0}^\infty \eta_{\ell^\prime} \hat{a}_{\ell^\prime 1}
\right) \ ,
\end{equation}
\begin{equation}
a_{\ell 2} = \hat{a}_{\ell 2} + \epsilon_{\ell} \left( \epsilon -
\sum_{\ell^\prime=0}^\infty \epsilon_{\ell^\prime} \hat{a}_{\ell^\prime 2}
\right) \ ,
\end{equation}
\begin{equation}
a_{\ell m} = \hat{a}_{\ell m} \quad (m > 2) \ .
\label{eqn:alm_inv}
\end{equation}
\end{subequations}
For simplicity, we have assumed that these equations are valid for all
$\ell$ with the prescription that the pivot coefficients
$\hat{a}_{\ell_\nu 0}$, $\hat{a}_{\ell_\kappa 0}$, $\hat{a}_{\ell_\eta
  1}$, $\hat{a}_{\ell_\epsilon 2}$ are zero. The matrix $\Sigma$ in
eq.~\eqref{eqn:al0_inv} is the covariance matrix between $\nu$ and
$\kappa$. As the peak variables are uncorrelated with the
$\hat{a}_{\ell m}$ coefficients, it is possible to put constraints in
$\nu$, $\kappa$, $\eta$ and $\epsilon$ without affecting the rest of
the degrees of freedom of the temperature, given by the variables
$\hat{a}_{\ell m}$. Once the peak constraints are imposed, the
original $a_{\ell m}$ coefficients are recovered using the inverse
relation. Notice that this inversion process is analytical and
therefore no numerical inversion is needed.

In addition to the temperature field, we can also consider the $E$ and
$B$ polarization fields. Although the peak selection is still done in
$T$, the $E$ and $B$-modes will be affected due to the corresponding
correlation between both fields and $T$. Once we have specified the
peak conditions on the temperature, we need to know what is the
conditional probability of $E$ and $B$, in order to calculate their
statistical properties. Although the primordial fluctuations do not
introduce correlation between the $B$-mode and the scalar fields,
there could be different physical effects which break the parity
invariance of the field and lead to the $TB$ and $EB$ correlations
\cite{contaldi2015,galaverni2015}. Within the formalism
established in this paper, we consider the general case where these
correlations are non-zero. If the distribution of the temperature and
polarization fields is Gaussian, then the conditional probability of
$E$ and $B$ given $T$ is a bivariate Gaussian with the following mean
values and covariance:
\begin{subequations}
\label{eqn:elm_blm_cov_peak}
\begin{equation}
\langle e_{\ell m} \rangle =
\frac{C_{\ell}^{TE}}{\sqrt{C_{\ell}^{TT}}} a_{\ell m} \ , \qquad
\langle b_{\ell m} \rangle =
\frac{C_{\ell}^{TB}}{\sqrt{C_{\ell}^{TT}}} a_{\ell m} \ ,
\label{eqn:elm}
\end{equation}
\begin{equation}
\mathbf{C} = \left( \begin{array}{cc}
C_\ell^{EE} & C_\ell^{EB} \\
C_\ell^{EB} & C_\ell^{BB} \\
\end{array} \right) - \frac{1}{C_\ell^{TT}} \left( \begin{array}{cc}
\left( C_\ell^{TE} \right)^2 & C_\ell^{TE} C_\ell^{TB} \\ C_\ell^{TE}
C_\ell^{TB} & \left( C_\ell^{TE} \right)^2 \\
\end{array} \right) \ ,
\label{eqn:elm_cov}
\end{equation}
\end{subequations}
where $e_{\ell m}$ and $b_{\ell m}$ are the spherical harmonic
coefficients of $E$ and $B$ respectively. The mean values of the
polarization modes are affected by the temperature field, which is
described by the $a_{\ell m}$ coefficients in these equations. The
constraints on the temperature due to the peak induce a non-zero
pattern in the polarization fields. This fact is used in
section~\ref{sec:multipolar} to calculate the shape of peaks in
polarization.

\section{Extrema statistics}
\label{sec:extrema_stat}

In this section we show how to select minima or maxima using the peak
variables $\nu$, $\kappa$, $\eta$ and $\epsilon$. In order to have a
critical point, the only requirement is to fix the first derivatives
to zero, that is, $\eta = 0$. In addition, if we want to have an
extremum, additional constraints in the mean curvature $\kappa$ and in
the eccentricity $\epsilon$ are needed. In particular, we ensure that
the critical point is an extremum by requiring that the eigenvalues of
the Hessian matrix have the same sign. This is done by imposing that
$|\epsilon| \leq \sqrt{a}|\kappa|$, where $a = \sigma_\kappa^2 /
\sigma_\epsilon^2$ is the ratio of the variance of the Laplacian
$\nabla^2T$ and that of $(\slashed{\partial}^*)^2T$. Whether that
extremum is a minimum or maximum depends on the sign of the
curvature. If $\kappa > 0$, the field will have a maximum, and, if
$\kappa < 0$, a minimum (in the case of $\kappa = 0$, the point would
be flat up to second order, but the probability of this is
zero). These extremum constraints can be imposed by considering the
probability of the peak degrees of freedom:
\begin{equation}
P(\nu,\kappa,\epsilon) \ \mathrm{d}\nu \ \mathrm{d}\kappa
\ \mathrm{d}^2\epsilon = \frac{2|\epsilon|}{2\pi\sqrt{(1-\rho^2)}}
\exp \left[-\frac{\nu^2-2\rho\nu\kappa+\kappa^2}{2(1-\rho^2)}
-|\epsilon|^2 \right] \ \mathrm{d}\nu \ \mathrm{d}\kappa
\ \mathrm{d}|\epsilon| \ \frac{\mathrm{d}\alpha}{\pi} \ ,
\label{eqn:prob}
\end{equation}
where the eccentricity is given by $\epsilon = |\epsilon|e^{i2\alpha}$,
that is, $|\epsilon|$ is the modulus and $\alpha$ is the orientation
of the ellipse. Notice that the peak height and the curvature are not
independent, with a joint probability given by a multivariate
Gaussian, where $\rho$ is the correlation. It is possible to write the
correlation as a function of the field variances: $\rho =
\sigma_\eta^2 / \sigma_\nu \sigma_\kappa$. As the eccentricity is a
Gaussian complex number, its modulus follows the Rayleigh
distribution, and the orientation angle $\alpha$ is distributed
uniformly in the interval $[0,\pi]$.

\begin{figure}
\begin{center}
\includegraphics[scale=0.29]{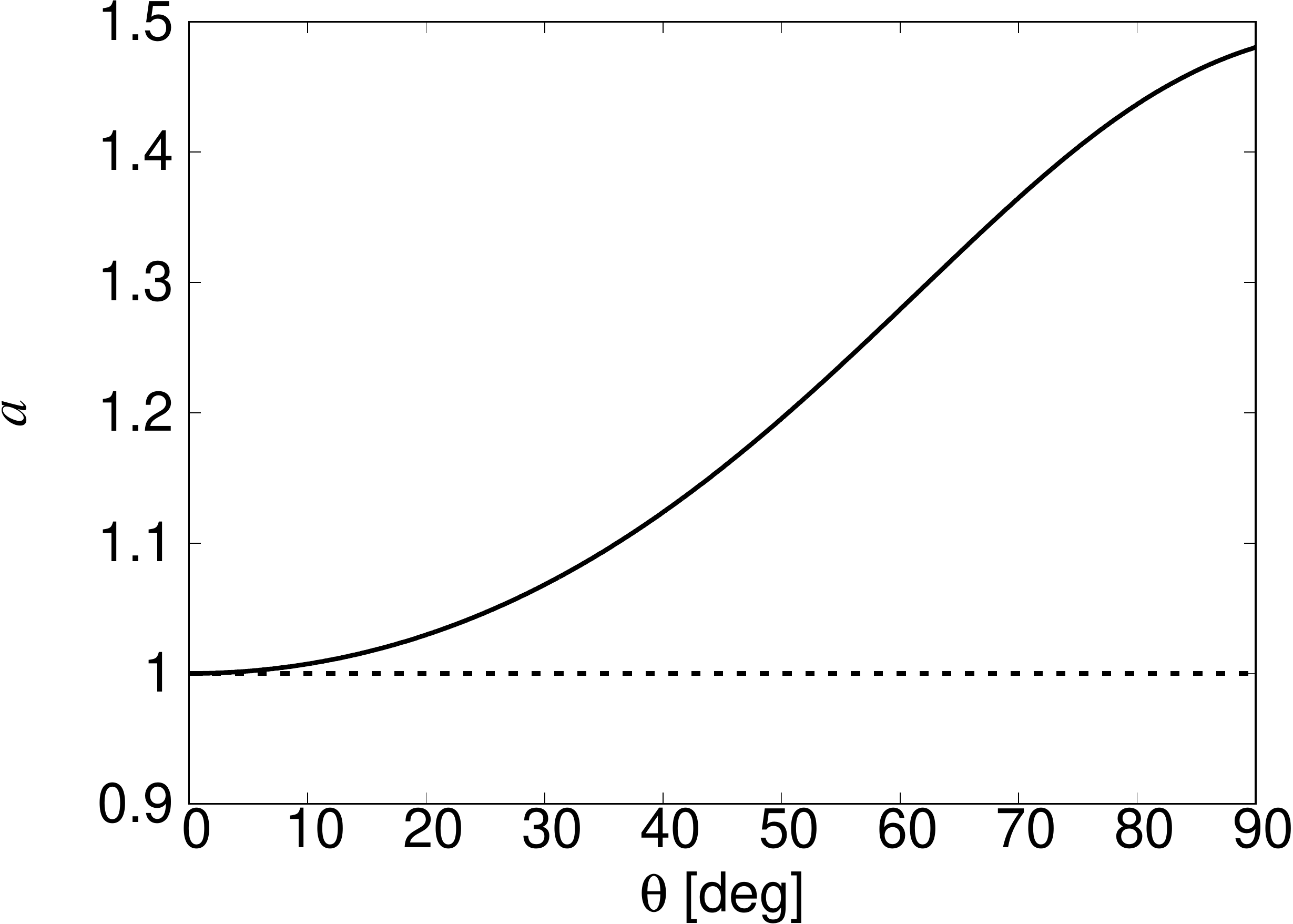}
\includegraphics[scale=0.29]{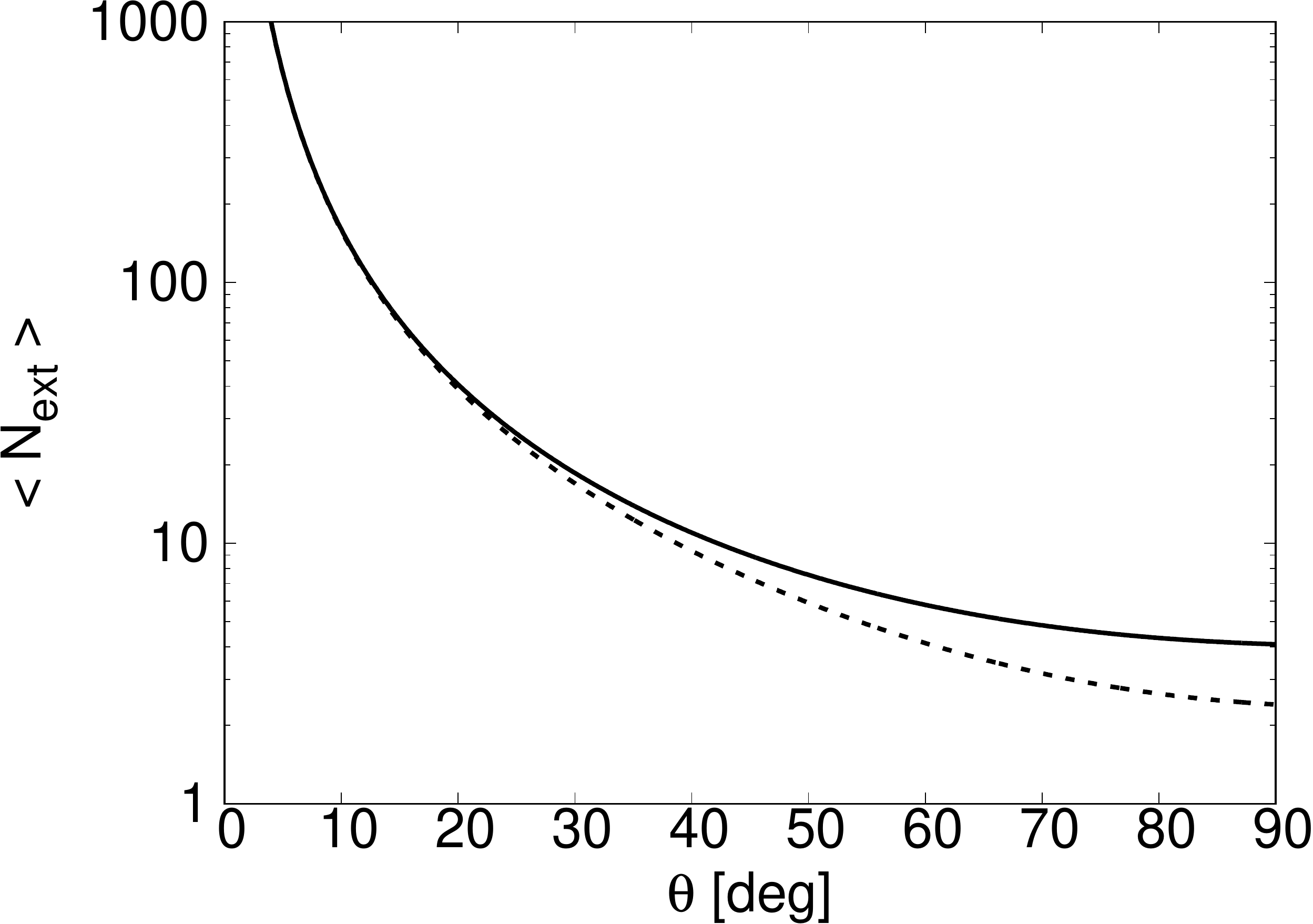}
\end{center}
\caption{The parameter $a = \sigma_\kappa^2 / \sigma_\epsilon^2$
    and the expected number of extrema are depicted as a function of
    the angluar size of the peak. The solid lines represent the
    calculation for the sphere, whereas the dashed lines correspond to
    the flat approximation. In order to select peaks with a given
    size, the temperature field is filtered by a Gaussian whose FWHM
    corresponds to the peak scale considered. Notice that in the case
    of large peaks, for which the flat approximation breaks, it is
    necessary to use the sterographic projection to relate the FWHM of
    the Gaussian and the angular size of the peak ($R = 2 \tan
    \theta/2$).}
\label{fig:a_th_next}
\end{figure}

It is possible to calculate the number density of peaks from the
probability density in eq.~\eqref{eqn:prob}. The density of peaks
depends on the particular size of the peak, in addition to its
probability. For instance, the number of big spots is suppressed
because they occupy an area larger than the small ones. Hence, it is
expected that small spots are more abundant than large ones. The spot
size dependence is introduced through the determinant of the Hessian
matrix, which is proportional to the inverse of the square of the spot
size. The number density of peaks is given by
\begin{equation}
n(\nu,\kappa,\epsilon) \ \mathrm{d}\nu \ \mathrm{d}\kappa
\ \mathrm{d}^2\epsilon = \frac{1}{2\pi\theta_*^2} \left( a\kappa^2 -
|\epsilon|^2 \right) P(\nu,\kappa,\epsilon) \ \mathrm{d}\nu
\ \mathrm{d}\kappa \ \mathrm{d}^2\epsilon \ ,
\label{eqn:density}
\end{equation}
where $\theta_*^2 = 2\sigma_\eta^2/\sigma_\epsilon^2$. This expression
differs from the one in \cite{bond1987} in the $a$ parameter, but we
recover it when $a \approx 1$. It is possible to show that there
exists the constraint relation $a=1+\theta_*^2$ between both
parameters. In the small-scale limit, it is possible to consider that
$\theta_* \ll 1$, and then $a \approx 1$. Nevertheless, this limit is
not valid if the sample is dominated by large spots ( see
figure~\ref{fig:a_th_next}).

The expected total number of extrema on the sphere is obtained by
integrating eq.~\eqref{eqn:density} over all possible values of $\nu$
and $\kappa$. However, the integration over the eccentricity
$\epsilon$ must be done in the region where $|\epsilon| \leq \sqrt{a}
|\kappa|$, in order to guarantee that the point is an extremum. The
expected total number of extrema is
\begin{equation}
\langle N_{\text{ext}} \rangle = 2 \left( 1 +
\frac{1}{\theta_*^2\sqrt{3+2\theta_*^2}} \right) \ .
\label{eqn:next}
\end{equation}
This number only depends on the value of $\theta_*$. When $\theta_*$
is small, the number of extrema is proportional to $\theta_*^{-2}$. In
this case, we obtain the result in \cite{bond1987}. But we need to
consider the exact formula when the number of peaks is small, or
equivalently, when the field is dominated by large peaks. In
figure~\ref{fig:a_th_next}, the expected number of extrema is
represented as a function of the peak size, where it is possible to
see that the flat approximation breaks for peaks larger than
$30^\circ$. In practice, there are only several of such large peaks on
the sphere, and therefore the cosmic variance is not significantly
reduced by performing a stacking analysis. However, the study of
particularly large peaks is still useful to test the properties of the
CMB at large scales.

As concrete examples of eq.~\ref{eqn:next}, it is possible to check
this expression analytically for dipolar and quadrupolar patterns. In
the case of the dipole, we assume that $C_1^{TT} \neq 0$ and
$C_\ell^{TT} = 0$ for all $\ell \neq 1$. Therefore, the field will
have a dipole with random orientation and amplitude. The small-scale
limit cannot be taken in this case, since $\theta_*^2 = \infty$. The
number of extrema in a dipolar pattern is always $2$, independently of
the randomness of the field. This fact agrees with the prediction from
eq.~\eqref{eqn:next} for a random dipole, which is $\langle
N_\text{ext} \rangle = 2$. Repeating the same reasoning for a random
quadrupole ($C_2^{TT} \neq 0$ and $C_\ell^{TT} = 0$, for $\ell \neq
2$), we find that $\langle N_\text{ext} \rangle = 4 $, as expected
from a quadrupolar pattern which always has $2$ maxima and $2$
minima. In general, from eq.~\eqref{eqn:next} follows that $\langle
N_{\text{ext}} \rangle \geq 2$, reflecting the fact that there will be
always one minimum and one maximum in the sphere at least. This is a
consequence of the extreme value theorem applied to the sphere.

We finish this section commenting that there are two ways of assigning
probabilities to the peak variables, depending on the physical problem
we are addressing. If we are interested in studying the statistical
properties of a single peak on the sphere, then we have to use the
probability in eq.~\eqref{eqn:prob}. This probability gives the
distribution of the peak variables in a single point, independently of
any other location on the sphere. On the other hand, sometimes it is
useful to sum over a given population of peaks in order to enhance the
signal we want to measure. The distribution of the peak variables in
this stacking-like procedure is different from the one-point
distribution. In this case, the correct way to assign probabilities is
given by the number density in eq.~\eqref{eqn:density}. The peak
variables of the stacked points are distributed on the sphere
following the number density, instead of the one-point probability of
the peak variables.

\section{Multipolar profiles}
\label{sec:multipolar}

The expected 2-dimensional shape of peaks on the sphere depends on how
the peak variables are constrained. Indeed, if the peak degrees of
freedom are randomly distributed without any additional constraint,
then the expected pattern is zero. It is possible to see the peak
shape as an effect of a bias in the peak variables. For instance, if
we impose a threshold for the peak height $\nu$, then the randomness
of the field is broken and the value of $\langle \nu \rangle$ is
different from zero. This bias in the expected value of the peak
height generates a non-trivial pattern on the sphere. In this section
we only consider the peak height $\nu$, the mean curvature $\kappa$
and the eccentricity $\epsilon$ as the peak degrees of freedom,
because the first order derivatives are fixed to zero ($\eta=0$) by
definition of a peak.

The fact that the expected value of the eccentricity $\langle \epsilon
\rangle$ could be biased introduces a $\phi$ dependence in the peak
pattern on the sphere. In order to take into account these angular
dependence, we expand a generic field on the sphere $X(\theta,\phi)$
in the following way:
\begin{equation}
X(\theta,\phi) = \sum_{m=-\infty}^\infty X_m(\theta) \ e^{im\phi} \ .
\label{eqn:temp_multipolar}
\end{equation}
The profiles $X_m(\theta)$ represent contributions to the peak with
different rotational symmetry. Since the field $X$ is real, it is
satisfied that $X_m(\theta)^* = X_{-m}(\theta)$, where the number $m$
is the spin of the profile. The fact that peaks are determined through
their derivatives up to second order implies that the profiles with
spin $m>2$ vanish. The dipolar profile with $m=1$ is also zero because
the first derivatives are zero by definition. Only the scalar ($m=0$)
and quadrupolar ($m=2$) profiles contribute to this expansion.

The inverse transform of eq.~\eqref{eqn:temp_multipolar} is
\begin{equation}
X_m(\theta) = \frac{1}{2\pi} \int \mathrm{d}\phi
\ X(\theta,\phi) \ e^{-im\phi} \ .
\end{equation}
In particular, the scalar profile $X_0(\theta)$ is the $\phi$-average
of the field, that is, the standard profile when spherical symmetry is
assumed. The quadrupolar profile $X_2(\theta)$ is a correction term
due to the asymmetry introduced by the eccentricity of the peak. It is
useful to write the multipolar profiles by using the associated
Legendre polynomials:
\begin{equation}
X_m(\theta) = \sum_{\ell=m}^\infty \sqrt{\frac{2\ell+1}{4\pi}}
\sqrt{\frac{(\ell-m)!}{(\ell+m)!}} \ a_{\ell m}^{X} \ P_\ell^m(\cos\theta)
\ .
\label{eqn:temp_m}
\end{equation}
In the particular case when the peak is located in the north pole, the
coefficients $a_{\ell m}^X$ in this expansion coincide with the
spherical harmonics coefficients of the field $X$.

\subsection{Profiles in harmonic space}

In this subsection, the multipolar profiles for the CMB temperature
and polarization are calculated in harmonic space. For simplicity, it
is convenient to use the Stokes parameters in polar coordinates with
the origin at the centre of the peak. These polar parameters $Q_r$ and
$U_r$ are a rotated version of the standard $Q$ and $U$ ones (see
\cite{kamionkowski1997}). The $Q_r$ field represents radial or
tangential polarization patterns around the spot. If the sign of $Q_r$
is positive the polarization is radial, and tangential in the case in
which $Q_r$ is negative. On the other hand, $U_r$ represents the
polarization rotated $45^\circ$ with respect to $Q_r$, as in the
standard Stokes parameters. If the peaks are not oriented, then a
polarization field with rotational symmetry is expected, and
therefore, $Q_r$ and $U_r$ will not depend on $\phi$. This is not the
case when the polarization field is described using the Cartesian
Stokes parameters (examples of the expected patterns in this
particular case can be seen in \cite{komatsu2011} and
\cite{planck162015}). The azimuthal dependence introduced in this way
is due to the inappropriate choice of the coordinate system, and it
does not reflect the rotational symmetry of the polarization field.

The expected value of the multipolar profiles is calculated from
eqs.~\eqref{eqn:alms_inv} and eq.~\eqref{eqn:temp_m},
taking into account that $\langle \hat{a}_{\ell m} \rangle = 0$. We
also assume that the first derivatives are zero ($\langle \eta \rangle
=0$). Therefore, the multipolar profiles $\langle T_m(\theta) \rangle$
only depend on the average of the peak height, mean curvature and
eccentricity. Depending on how these mean values are constrained,
different shapes are obtained. In order to have a peak, the condition
imposed on the expected values of $\kappa$ and $\epsilon$ is the
extremum constraint ($|\epsilon| \leq \sqrt{a}|\kappa|$), which
guarantees to have a maximum or minimum. In general, there is more
freedom in choosing the value of $\nu$. For instance, if we are
interested in peaks above a given threshold $\nu_t$, then its expected
value must be calculated with the condition that $\nu >
\nu_t$. Another possibility is to fix $\nu$ to a given value and study
the pattern induced by the peak with that particular height. Since in
this paper we are interested in the qualitative behaviour of peaks,
this latter approach is used in the calculations.

Additionally, the polarization field of the peak in terms of $Q_r$ and
$U_r$ is calculated from the $E$ and $B$ modes. Notice that, if the
peak is located at the north pole, then $Q_r$ and $U_r$ coincide with
the standard Stokes parameters on the sphere. Therefore, they can be
calculated using its expansion in terms of the spin-weighted spherical
harmonics.

Firstly, we consider peaks with rotational symmetry. In this case, the
only expected contribution comes from the $m=0$ profile in the
multipolar expansion (eq.~\eqref{eqn:temp_multipolar}). The monopolar
profiles are therefore given by
\begin{subequations}
\begin{equation}
\langle T_0(\theta) \rangle = \sum_{\ell = 0}^{\infty} \frac{2\ell+1}{4\pi}
\left[ b_\nu + b_\kappa \ell(\ell+1) \right] C_\ell^{TT}
\ P_{\ell}(\cos\theta) \ ,
\end{equation}
\begin{equation}
\langle Q_{r0}(\theta) \rangle = - \sum_{\ell = 2}^{\infty}
\frac{2\ell+1}{4\pi} \sqrt{\frac{(\ell-2)!}{(\ell+2)!}}  \left[ b_\nu
  + b_\kappa \ell(\ell+1) \right] C_\ell^{TE} \ P_{\ell}^2(\cos\theta)
\ ,
\end{equation}
\begin{equation}
\langle U_{r0}(\theta) \rangle = - \sum_{\ell = 2}^{\infty}
\frac{2\ell+1}{4\pi} \sqrt{\frac{(\ell-2)!}{(\ell+2)!}}  \left[ b_\nu
  + b_\kappa \ell(\ell+1) \right] C_\ell^{TB} \ P_{\ell}^2(\cos\theta)
\ ,
\end{equation}
\end{subequations}
These profiles depend on the bias parameters $b_\nu$ and $b_\kappa$,
which can be calculated from the expected value of $\nu$ and $\kappa$:
\begin{equation}
\left( \begin{array}{c}
 b_\nu \sigma_\nu \\ b_\kappa \sigma_\kappa
\end{array} \right) = \Sigma^{-1}
\left( \begin{array}{c}
\langle \nu \rangle \\ \langle \kappa \rangle
\end{array} \right) \ .
\end{equation}
The matrix $\Sigma$ relating these quantities is the covariance matrix
of $\nu$ and $\kappa$, described in Section \ref{sec:uncorrelating}
(see also eq.~\eqref{eqn:sigma}). These profiles obtained for
spherical symmetric peaks represent the generalization of the
expressions in \cite{komatsu2011} for large angular distances (see
appendix~\ref{app:flat_approx}). As it is expected, the temperature
profile depends on the angular power spectrum $C_\ell^{TT}$. On the
other hand, the coefficients in the multipolar expansion of the Stokes
parameters are given by the cross-correlation between the temperature
and the polarization fields. In the case of a spherically symmetric
peak, it is possible to see that $Q_{r0}$ describes the gradient of
the polarization field, while $U_{r0}$ represents the curl
contribution. For this reason, $Q_{r0}$ depends exclusively on the
correlation of the temperature with $E$, which is the gradient of the
polarization field, and $U_{r0}$ depends on the correlation of $T$
with $B$, which is the curl contribution. If it is assumed that there
is not any physical effect capable of rotating the polarization angle
(e.g., birefringence \cite{galaverni2015}) or violating parity
conservation \cite{contaldi2015} then the correlation between $T$ and
$B$ vanishes, and hence the expected value of $U_{r0}$ is zero.

\begin{figure}
\begin{center}
\includegraphics[scale=0.55]{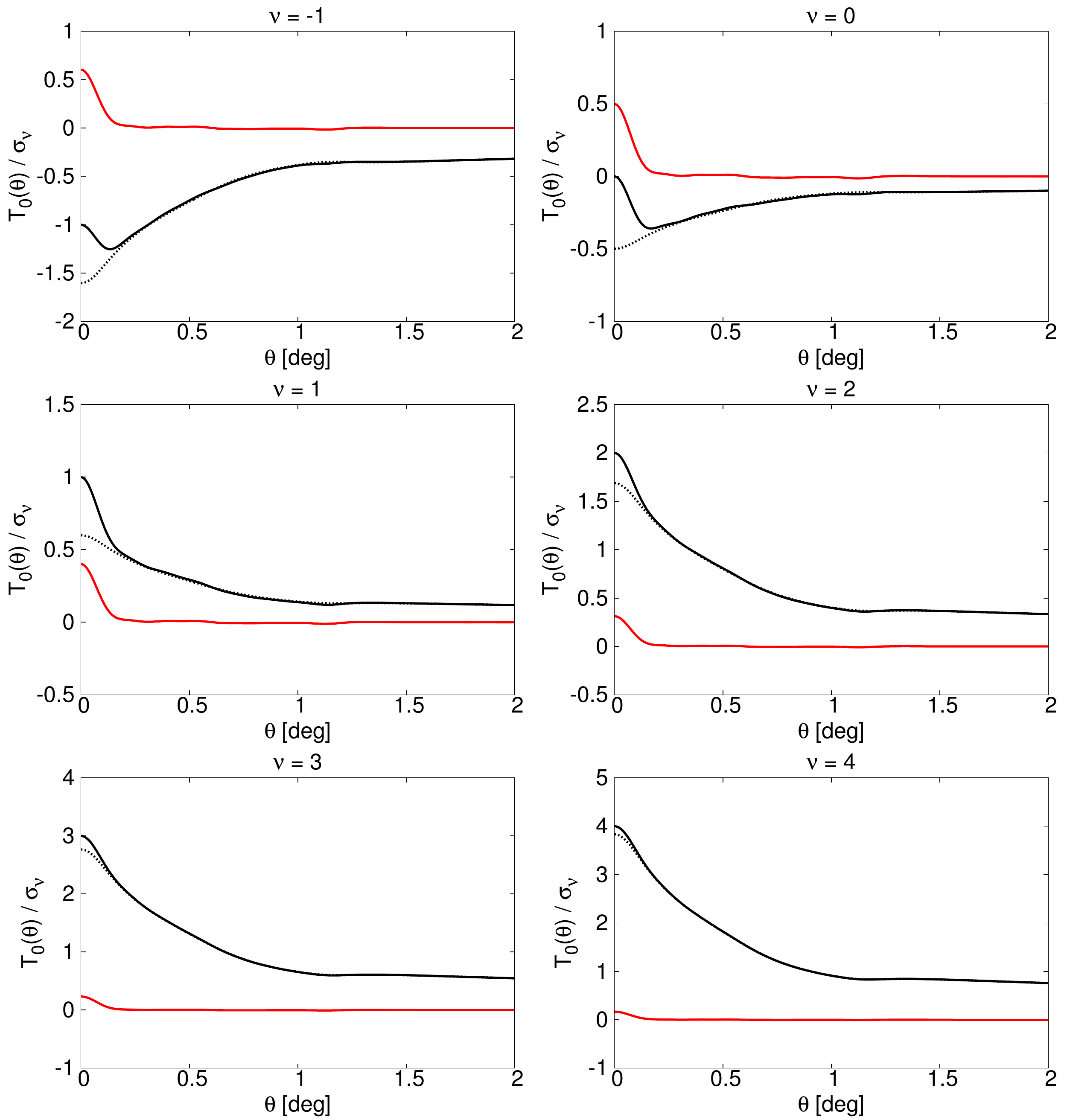}
\end{center}
\caption{Spherically symmetric temperature profiles $T_0(\theta)$ for
  maxima conditioned to different peak heights. The dotted black lines
  depicts the contribution proportional to the temperature correlation
  function (peak height bias $b_\nu$), while the red line corresponds
  to the modification due to its Laplacian (mean curvature bias
  $b_\kappa$). The black solid line represents the total profile.}
\label{fig:profile_t0}
\end{figure}

\begin{figure}
\begin{center}
\includegraphics[scale=0.55]{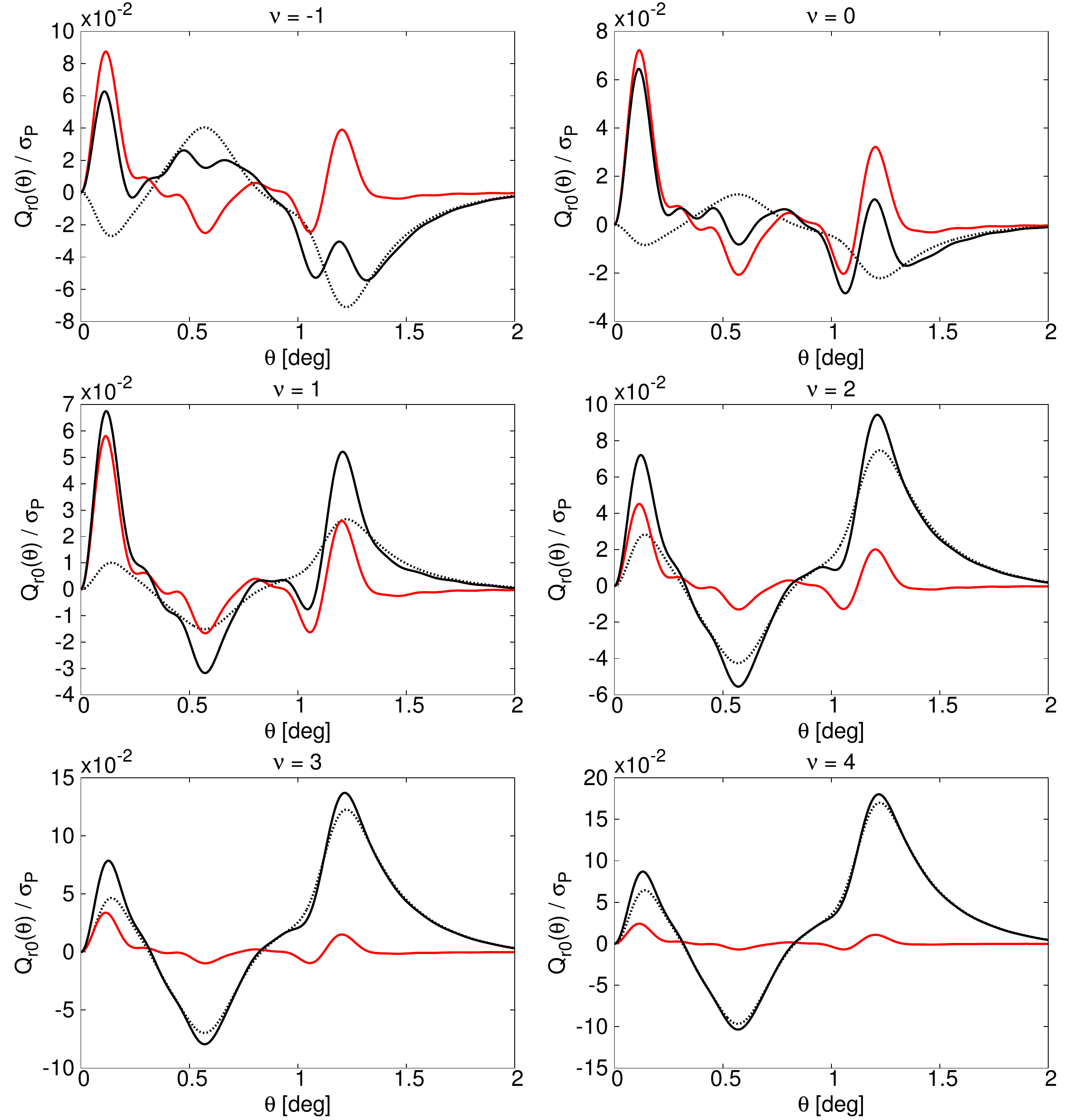}
\end{center}
\caption{Monopolar profiles $Q_{r0}(\theta)$ for different peak
  heights. The dotted black lines show the contribution due to the
  peak height bias ($b_\nu$), while the red line is the modification
  caused by the mean curvature bias ($b_\kappa$). The black solid line
  corresponds to the total profile. All these profiles are normalized
  by $\sigma_P = \sqrt{\langle Q^2\rangle + \langle U^2 \rangle}$.}
\label{fig:profile_qr0}
\end{figure}

The monopolar temperature profiles for maxima and different values of
the peak height are represented in figure~\ref{fig:profile_t0},
showing the two effects due to the peak height and curvature
biases. The curvature term contributes to the peak only at small
scales, just modifying the peakedness of the profile. For large values
of $\nu$, the monopolar peak profile tends to be proportional to the
temperature correlation function, since the curvature bias becomes
negligible (see figure~\ref{fig:biases} and the discussion in
section~\ref{subsec:bias} about the behaviour of the bias
parameters). In the case of polarization, we consider only $Q_{r0}$
because the $TB$ power spectrum is zero in the standard model. The
monopolar profiles of $Q_{r}$ for maxima, conditioned to the value of
$\nu$, are represented in figure~\ref{fig:profile_qr0}. As in the
temperature case, the contribution to the $Q_{r0}$ profile for high
$\nu$ comes from the correlation between the temperature and
$Q_r$. The effect due to the curvature bias, present in profiles with
small $\nu$, tends to modify the peaks of the $Q_{r0}$ profile. The
profiles represented in figures~\ref{fig:profile_t0} and
\ref{fig:profile_qr0} are calculated for maxima, but equivalent
results are obtained for minima.

In the case that $\langle \epsilon \rangle \neq 0$ (e.g., when the
peaks are oriented towards some direction), then there is also a
contribution to the quadrupolar profile ($m=2$) in
eq.~\eqref{eqn:temp_multipolar}:
\begin{subequations}
\label{eqn:x2}
\begin{equation}
\langle T_2(\theta) \rangle = b_\epsilon \sum_{\ell = 0}^\infty
\frac{2\ell+1}{4\pi} C_\ell^{TT} \ P_\ell^2(\cos \theta) \ ,
\label{eqn:t2}
\end{equation}
\begin{equation}
\langle Q_{r2}(\theta) \rangle = -2 b_\epsilon \sum_{\ell = 0}^\infty
\frac{2\ell+1}{4\pi} \sqrt{\frac{(\ell-2)!}{(\ell+2)!}} \left[ C_\ell^{TE}
\ P_\ell^{+}(\cos \theta) + i C_\ell^{TB} \ P_\ell^{-}(\cos \theta)
\right] \ ,
\label{eqn:qr2}
\end{equation}
\begin{equation}
\langle U_{r2}(\theta) \rangle = 2 i b_\epsilon \sum_{\ell = 0}^\infty
\frac{2\ell+1}{4\pi} \sqrt{\frac{(\ell-2)!}{(\ell+2)!}} \left[ C_\ell^{TE}
\ P_\ell^{-}(\cos \theta) + i C_\ell^{TB} \ P_\ell^{+}(\cos \theta)
\right] \ .
\label{eqn:ur2}
\end{equation}
\end{subequations}
In this case, the bias of the quadrupolar profiles is defined as
$b_\epsilon = \langle \epsilon \rangle / \sigma_\epsilon$, which is
proportional to the expected value of the eccentricity. The $\theta$
dependence in eqs.~\eqref{eqn:qr2} and \eqref{eqn:ur2} is described by
the functions:
\begin{subequations}
\label{eqn:pl+_pl-}
\begin{equation}
P_\ell^{+}(x) = - \left[ \frac{\ell-4}{1-x^2} + \frac{1}{2} \ell
  \left(\ell-1\right) \right] P_\ell^2(x) + \left( \ell + 2 \right)
\frac{x}{1-x^2} P_{\ell-1}^2(x) \ ,
\label{eqn:pl+}
\end{equation}
\begin{equation}
P_\ell^{-}(x) = - 2 \left[ \left( \ell - 1 \right) \frac{x}{1-x^2}
  P_\ell^2(x) - \left( \ell + 2 \right) \frac{1}{1-x^2}
  P_{\ell-1}^2(x) \right] \ .
\label{eqn:pl-}
\end{equation}
\end{subequations}
These functions arise in the analysis of any $2$-spin field on the
sphere (e.g., CMB polarization or weak lensing). They define the
$\theta$ dependence of the $2$-spin spherical harmonics with $m=2$ as
a function of
$\ell$ \cite{kamionkowski1997,stebbins1996}. These
expressions are undetermined in $\theta = 0$ ($x=1$), but they have a
continuous limit if the following values are adopted
(see \cite{stebbins1996}):
\begin{equation}
P_\ell^{\pm}(1) = \pm \frac{1}{4} \frac{(\ell+2)!}{(\ell-2)!}
\end{equation}

\begin{figure}
\begin{center}
\includegraphics[scale=0.28]{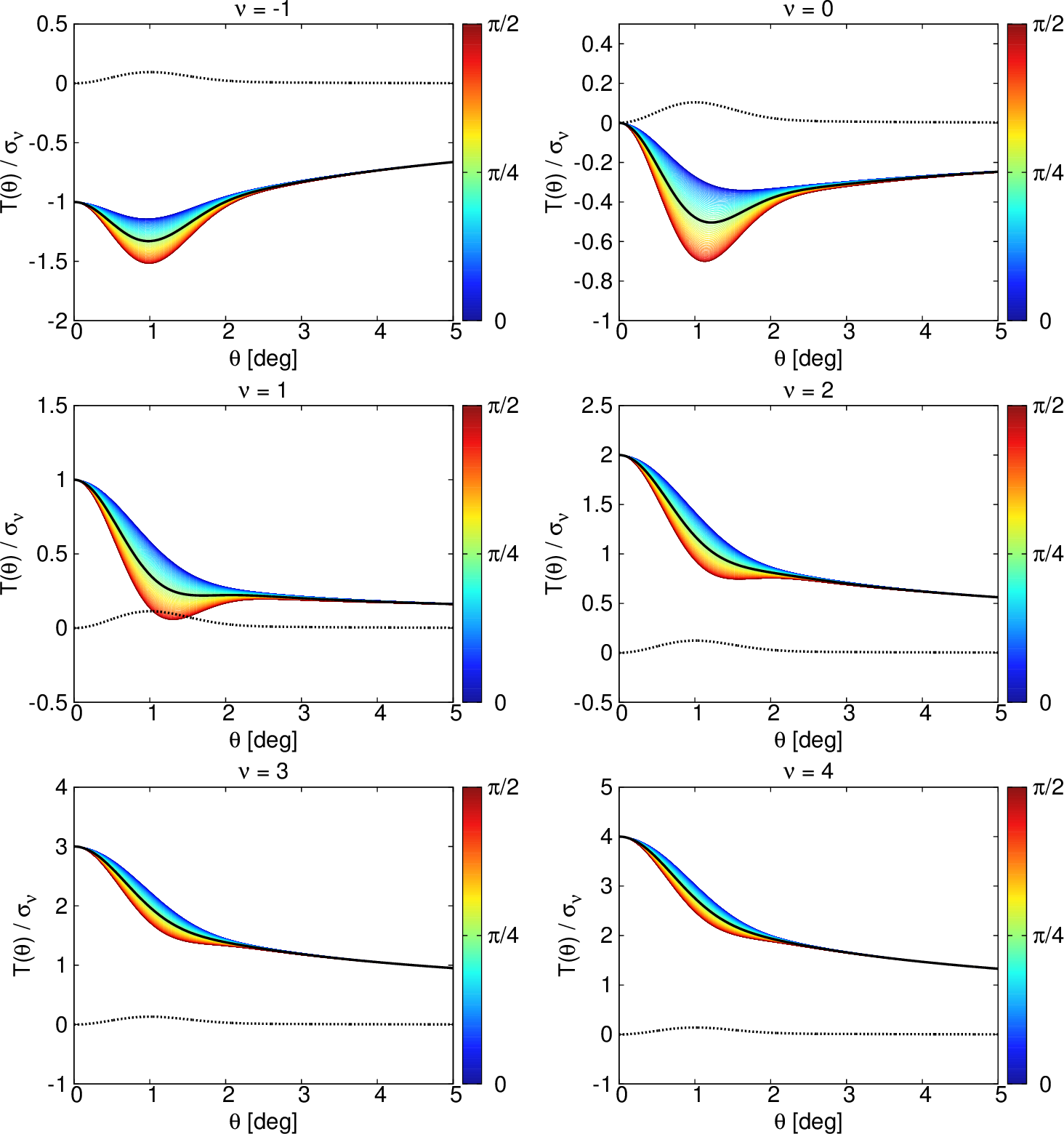}
\end{center}
\caption{The effect of the eccentricity on the temperature profile for
  different $\nu$. The principal axes of the peak are oriented
  according to the Cartesian system of reference, which implies that
  $\epsilon$ is real and the eccentricity bias is given by $b_\epsilon
  = \langle | \epsilon | \rangle / \sigma_\epsilon$.  The black solid
  line depicts the spherically symmetric profile ($m=0$). The color
  scale represents how the peak profile varies as a function of the
  azimuthal angle $\phi$. The maximum and minimum elongations are
  reached at $\phi=0$ and $\phi=\pi/2$ respectively. The quadrupolar
  profiles ($m=2$) are represented by black dotted lines. In this
  figure, the temperature field is filtered by a Gaussian of FWHM
  $1^\circ$.}
\label{fig:profile_t2}
\end{figure}

\begin{figure}
\begin{center}
\includegraphics[scale=0.28]{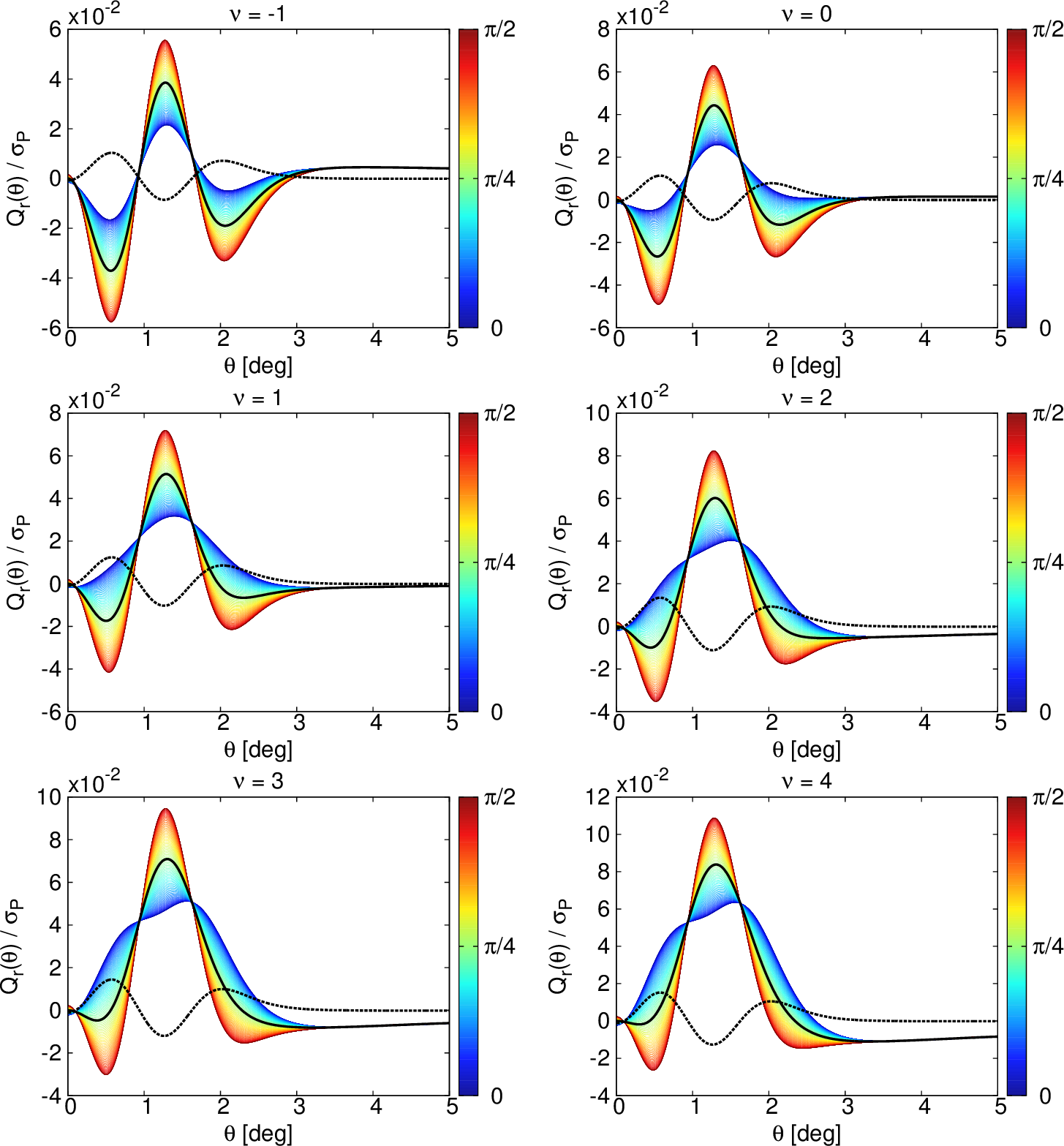}
\end{center}
\caption{The effect of the eccentricity on the $Q_r$ profile for
  different $\nu$. The peaks are oriented in the same way than in
    figure~\ref{fig:profile_t2}. The black solid line corresponds to
  the spherically symmetric profile ($m=0$). The color scale
  represents how the profile varies as a function of the azimuthal
  angle $\phi$. The maximum and minimum elongations are reached at
  $\phi=0$ and $\phi=\pi/2$ respectively. The quadrupolar profiles
  $Q_{r2}(\theta)$ are represented by black dotted lines. In this
  figure, the peak is selected in the temperature field, filtered by a
  Gaussian of FWHM $1^\circ$. All these profiles are normalized by
  $\sigma_P = \sqrt{\langle Q^2\rangle + \langle U^2 \rangle}$.}
\label{fig:profile_qr2}
\end{figure}

\begin{figure}
\begin{center}
\includegraphics[scale=0.50]{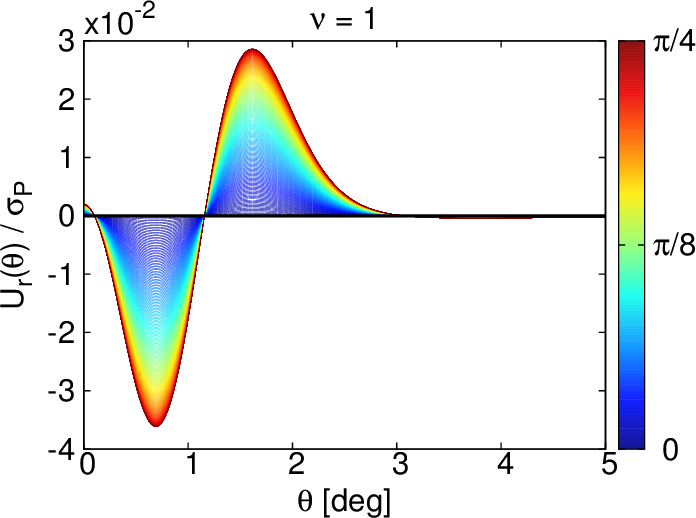}
\end{center}
\caption{The effect of the eccentricity on the $U_{r}$ profile for
  $\nu=1$. The peaks are oriented in the same way than in
  figure~\ref{fig:profile_t2}. The color scale represents how the
  profile varies as a function of the azimuthal angle $\phi$. This
  profile vanishes for $\phi=0 $ and it increases until reaching the
  maximum contribution at $\phi=\pi/4$. Different values of $\nu$ only
  change the amplitude of this profile following the dependence of $|
  b_\epsilon |$ as a function of the peak height (see
  figure~\ref{fig:biases}). In this figure, the peak is selected in
  the temperature field, filtered by a Gaussian of FWHM $1^\circ$. The
  profile is represented normalizing by $\sigma_P = \sqrt{\langle
    Q^2\rangle + \langle U^2 \rangle}$.}
\label{fig:profile_ur2}
\end{figure}

\begin{figure}
\begin{center}
\includegraphics[scale=0.45]{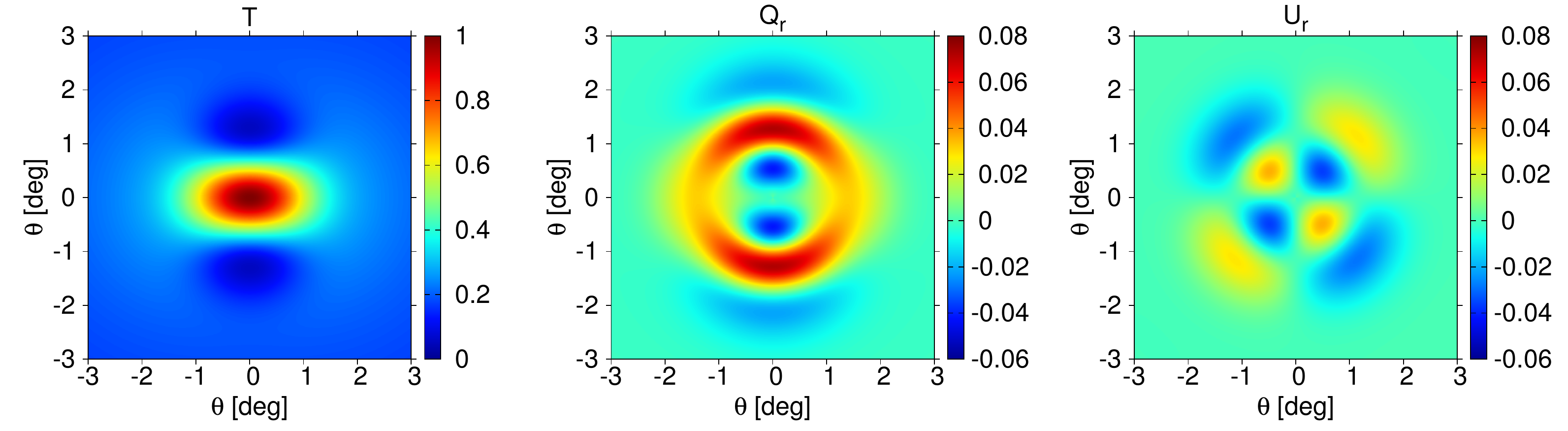}
\end{center}
\caption{The 2-dimensional shape of peaks with eccentricity for
    oriented peaks with $\nu = 1$. The panels from left to right
  represents $T$, $Q_r$ and $U_r$. In this figure, only the
  temperature field is filtered with a Gaussian of FWHM $1^\circ$. The
  units of color scales are given in terms of $\sigma_\nu$ for the
  temperature, and $\sigma_P$ for the Stokes parameters.}
\label{fig:patch}
\end{figure}

\begin{figure}
\begin{center}
\includegraphics[scale=0.65]{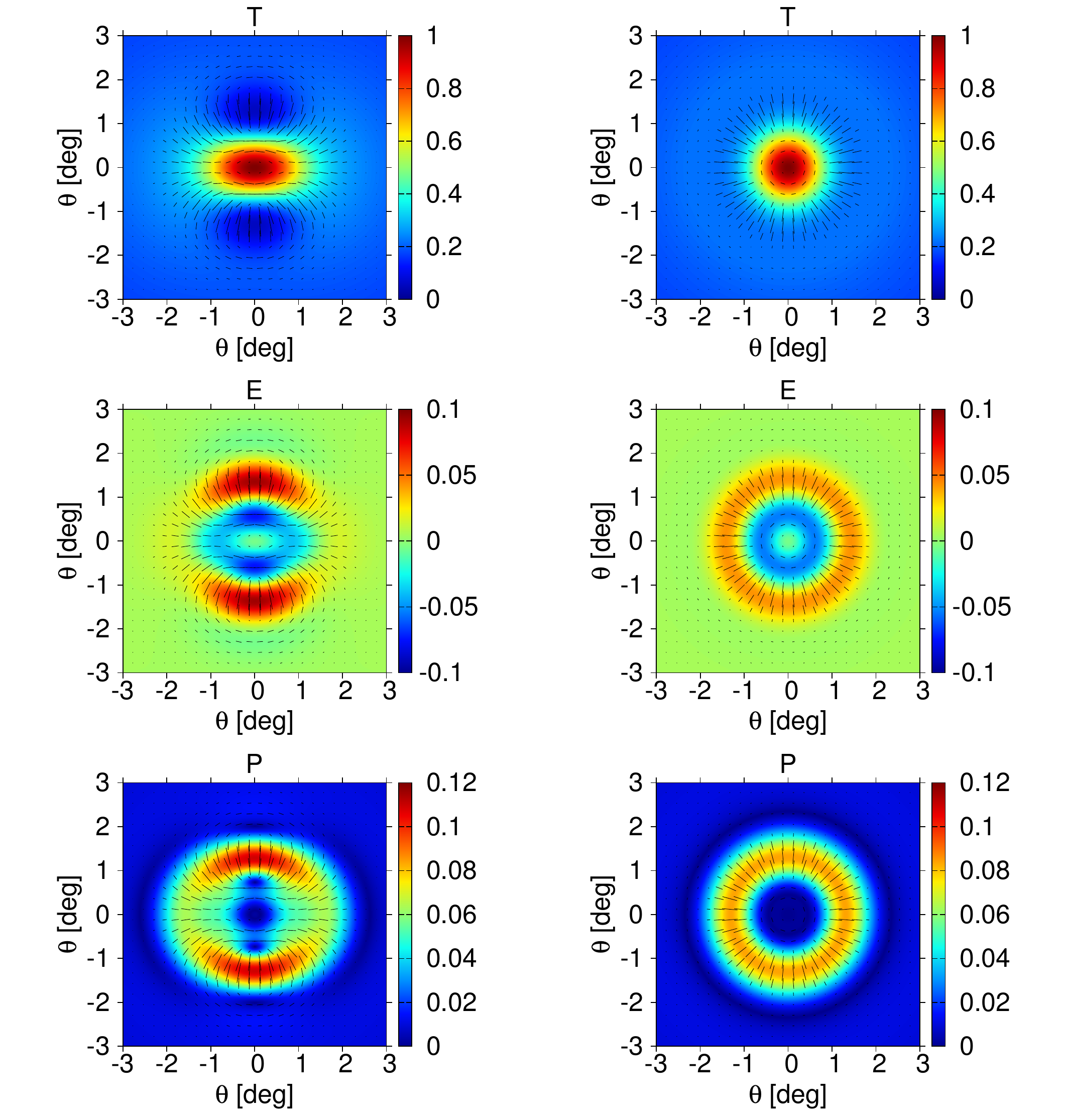}
\end{center}
\caption{The 2-dimensional shape of peaks for $\nu = 1$. In the left
  panels, it is considered a oriented peak with eccentricity,
  whilst in the right panels it is represented a spherical symmetric
  peak. In all these figures, the polarization directions are drawn
  over it. The length of the headless vectors is proportional to the
  polarization degree. \emph{Upper row}: the color map represents the
  temperature pattern induced by the peak.  \emph{Middle row}: in this
  case the color map depicts the $E$-mode polarization. \emph{Lower
    row}: it is represented $P \equiv \sqrt{Q^2+U^2}$, which describes
  the degree of polarization. In this figure, only the temperature
  field is filtered with a Gaussian of FWHM $1^\circ$. The units of
  color scales are given in terms of $\sigma_\nu$ for the temperature,
  and $\sigma_P$ for the $E$-mode and $P$.}
\label{fig:patch_pol}
\end{figure}

The quadrupolar profiles in eqs.~\eqref{eqn:x2} are complex quantities
whose phase represents a rotation of the system of reference. The
principal axes coincide with the $xy$ axes when the eccentricity bias
$b_\epsilon$ is real. Regarding the CMB polarization, one difference
with respect to the spherically symmetric case is that the
polarization fields $E$ and $B$ contribute to both Stokes parameters
$Q_r$ and $U_r$. The Stokes parameters in polar coordinates describe
properly peaks with rotational symmetry. However, when the peak has
nonzero eccentricity, the gradient and curl contributions are mixed
due to the elongation of the peak. The effect of the eccentricity bias
on the temperature and polarization peak shapes is represented in
figures~\ref{fig:profile_t2}-\ref{fig:profile_ur2} as a function of
the peak height. In these figures, the mean value of the eccentricity
has been calculated from the probability density distribution in
eq.~\eqref{eqn:prob}, imposing the condition that $\epsilon$ is real
($\alpha = 0$). Geometrically, this is equivalent to orient the peak,
such that the principal axes coincide with the Cartesian system of
reference. Additionally, the 2-dimensional shape of temperature peaks
and its effect on the Stokes parameters are shown in
figures~\ref{fig:patch} and \ref{fig:patch_pol}.

Let us remark that the multipolar profiles have already been used to
test the standard cosmological model with the Planck temperature and
polarization data \cite{planck162015}.\footnote{ In
  \cite{planck162015}, the multipolar profiles of polarization are
  defined expanding the quantity $Q+iU$, where the Stokes parameters
  are given in Cartesian coordinates (only valid in the flat
  approximation). The profiles $P_m$ arising in this expansion are
  related to the ones used in this work as follows:
\[ P_0 = Q_{r2} - i U_{r2} \ , \]
\[ P_2 = Q_{r0} \ , \]
\[ P_4 = Q_{r2} + i U_{r2} \ . \]
} The orientation of peaks in that work is performed by selecting the
principal axes in the inverse Laplacian of the temperature. This
allows to reduce the noise contribution in order to have a better
estimation of the orientation axes. On the other hand, the theoretical
calculations in the present paper are done directly in the temperature
field, but the formalism can be trivially generalized so that the peak
is selected and oriented in any derived field. For instance, it is
possible to select the peak in a smoothed version of the temperature
(in particular the inverse Laplacian), in order to reduce the noise or
study the physics of peaks at different scales. In this sense, the
work in this paper complements the study in \cite{planck162015} giving
a theoretical background, which is completely general and can be
applied to many situations.

\subsection{Profiles in real space}

In this subsection, we provide an alternative description of the peak
profiles, in which they are expressed in terms of derivatives of
different correlation functions, depending on which field is
considered and where the peak is selected. In the following, it is
assumed that the peak is located in the temperature field and its
effect on a general field $X$, which can be $T$, $E$, $B$, $Q_r$ or
$U_r$, is studied. It is straightforward to generalize this formalism
for peaks selected in any other field replacing $T$ by that
field. Using vector notation we have that $\langle X_m(\theta)
\rangle$ can be written as the following dot products:
\begin{subequations}
\label{eqn:temp_profile}
\begin{equation}
\langle X_0(\theta) \rangle = \mathbf{b}_0^\dagger
\ \mathbf{C}_0^{TX}(\theta) \ ,
\label{eqn:temp_profile_0}
\end{equation}
\begin{equation}
\langle X_2(\theta) \rangle = \mathbf{b}_2
\ \mathbf{C}_2^{TX}(\theta) \ ,
\label{eqn:temp_profile_2}
\end{equation}
\end{subequations}
where the biases $\mathbf{b}_0$ and $\mathbf{b}_2$ are defined as
\begin{equation}
\mathbf{b}_0 \equiv \left(
\begin{array}{c}
b_\nu \\ b_\kappa
\end{array}
\right) \ , \qquad \mathbf{b}_2 \equiv b_\epsilon \ .
\end{equation}
The biases concerning the scalar degrees of freedom $\nu$ and $\kappa$
are combined in the vector $\mathbf{b}_0$, while the bias related to
the eccentricity is denoted by $\mathbf{b}_2$ for convenience. The
$\theta$-dependence of the multipolar profiles is calculated from the
correlation function $C^{TX}(\theta)$:

\begin{equation}
\mathbf{C}_0^{TX}(\theta) =
\left( \begin{array}{c}
1 \\ - \nabla^2
\end{array} \right) C^{TX}(\theta) \ , \qquad
\mathbf{C}_2^{TX}(\theta) = (\slashed{\partial}^*)^2 C^{TX}(\theta) \ .
\label{eqn:c0_c2}
\end{equation}
The first component of the vector $\mathbf{C}_0^{TX}$ is the
correlation function itself, while the second one is minus its
Laplacian. On the other hand, the function $\mathbf{C}_2^{TX}$
defining the quadrupolar profile is written as a second order
covariant derivative of the correlation function.

The quantities defined in eq.~\eqref{eqn:c0_c2}, which determine the
shape of the peak, are different derivatives of the correlation
function. Indeed, these derivatives are the cross-correlations of the
field $X$ with the peak degrees of freedom. For instance, the
Laplacian of the correlation function $\nabla^2 C^{TX}(\theta)$ is
proportional to the correlation of the mean curvature $\kappa$ and the
field $X$, that is, $\langle \kappa X \rangle$. On the other hand, the
derivative $(\slashed{\partial}^*)^2 C^{TX}(\theta)$ is proportional
to the correlations $\langle \epsilon X \rangle$, while the
correlation function itself is proportional to $\langle \nu X
\rangle$. The fact that the field $X$ is correlated with the peak
degrees of freedom is the reason why any constraint on the peak
variables $\nu$, $\kappa$ and $\epsilon$ modifies the shape of the
peak.

\subsection{Bias discussion}
\label{subsec:bias}

The terms contributing to the multipolar profiles in
eqs.~\eqref{eqn:temp_profile} arise from different peak selection
biases. There are three conditions that can be imposed on peaks:
constraints on the peak height, the condition of being a maximum or
minimum and constraints on the orientation of the peak. The condition
of being an extremum affects to the mean curvature and the
eccentricity (see section~\ref{sec:extrema_stat}). Hereafter, the bias
parameters are calculated conditioning to the value of $\nu$.

\begin{figure}
\begin{center}
\includegraphics[scale=0.58]{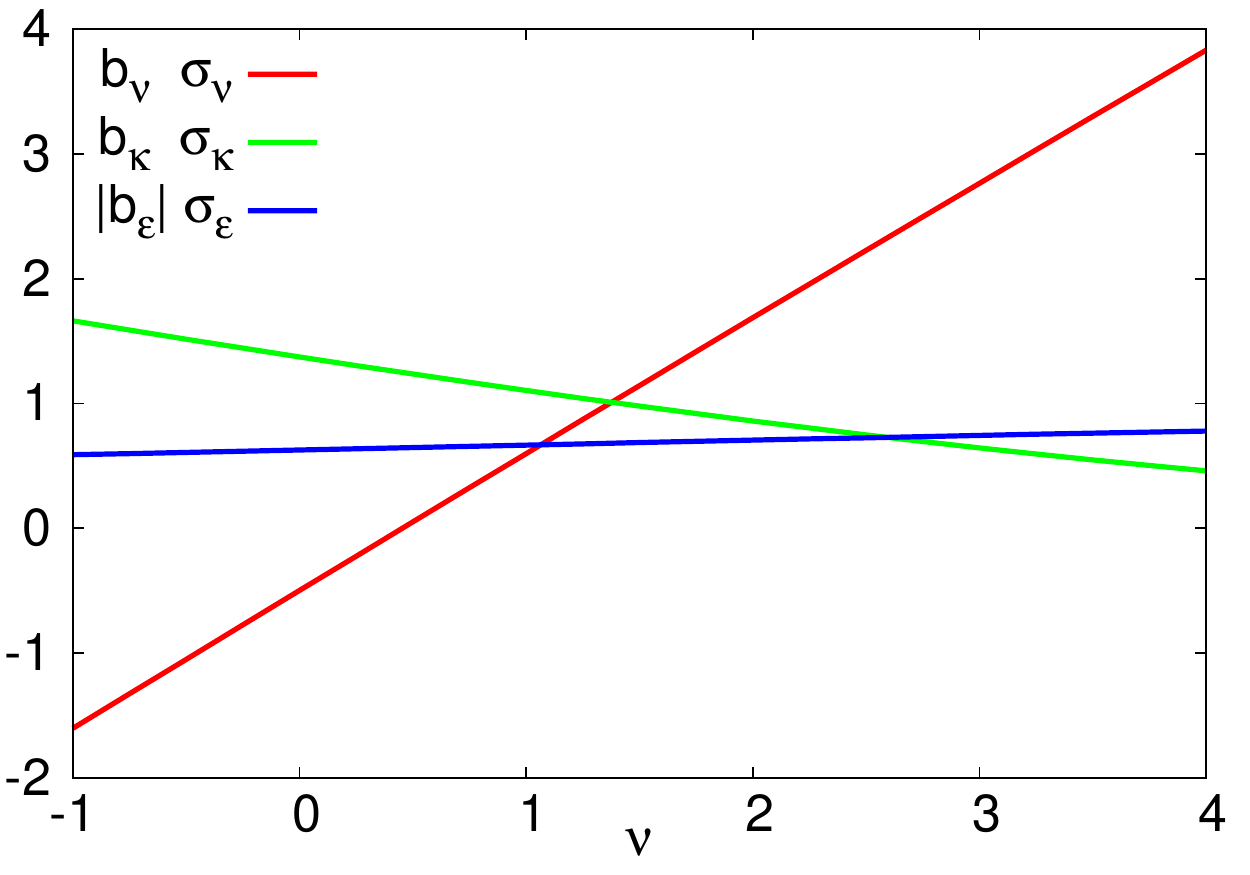}
\includegraphics[scale=0.58]{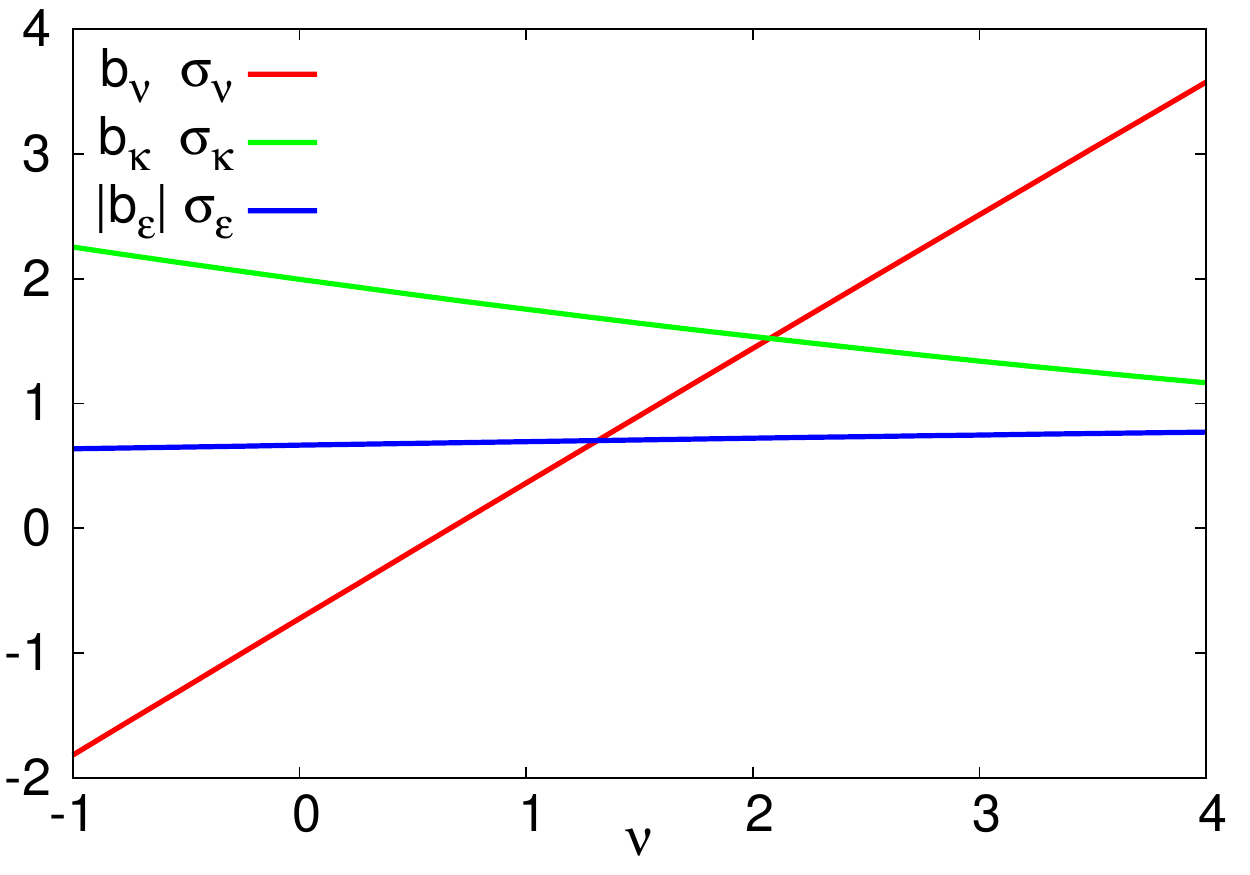}
\end{center}
\caption{Biases of the peak profile for maxima as a function of the
  peak height $\nu$. The curvature and the eccentricity are
  marginalised, while the peak height is conditioned to a given
  value. \emph{Left}: The one-point probability is used for averaging
  the peak variables and calculating the biases. \emph{Right}: The
  biases are calculated using the peak number density.}
\label{fig:biases}
\end{figure}

The biases for maxima are represented in figure~\ref{fig:biases} as a
function of the peak height. In the high- peak limit, the maximum
selection has no effect on the profile because it is more likely that
a peak with high $\nu$ is a maximum, without any additional bias on
the curvature. Therefore, the curvature bias $b_\kappa$ approaches to
zero for high $\nu$ (the expected value of the mean curvature is
$\langle \kappa \rangle \sim \rho \nu$ for large values of $\nu$). We
arrive at the same conclusion if we consider minima with extreme
negative values of $\nu$. On the other hand, the peak height bias
$b_\nu$ approaches to $\nu / \sigma_\nu$ in the high-peak limit (see
figure~\ref{fig:biases}). Hence, the radial profile of high peaks is
proportional to the correlation function.

Finally, we consider constraints on the eccentricity. If the peaks are
oriented according to its principal axes, then the mean value $\langle
\epsilon \rangle$ is not zero, introducing a bias in the value of
$\epsilon$. The quadrupolar profile ($m=2$) in
eq.~\eqref{eqn:temp_profile_2}, which breaks the rotational symmetry
and introduces an azimuthal dependence in the peak shape, is
proportional to the bias $b_\epsilon = \langle \epsilon \rangle /
\sigma_\epsilon$. As this bias is a complex number whose argument only
has information about the orientation angle, then the statistical
properties of the eccentricity are only in its modulus $| b_\epsilon
|$. In the high-peak limit, the modulus of the eccentricity bias
approaches to $| b_\epsilon | = \sqrt{\pi}/2\sigma_\epsilon$.

In figure~\ref{fig:biases}, we consider two different ways of
calculating the biases. In one case, the one-point probability
(eq.~\eqref{eqn:prob}) is used for averaging the peak variables and,
in the other case, it is used the number density of peaks
(eq.~\eqref{eqn:density}). Each of these approaches are useful in
different situations (see Section \ref{sec:extrema_stat} for a
discussion). Although the biases must be independent of the
probability used for their calculation in the high-peak limit,
differences can be seen for large values of $\nu$.

The eccentricity is a complex number whose phase describes the
orientation of the peak. It is possible to remove this phase choosing
the principal axes of the peak as the reference system. In this
particular case, both the eccentricity $\epsilon$ and the bias
$b_\epsilon$ are real. Combining eqs.~\eqref{eqn:temp_profile}, it can
be shown that the expected value $\langle T(\theta,\phi) \rangle$ is a
biased correlation function, where the bias can be seen as the
operator
\begin{equation}
b = b_\nu - \left( b_\kappa - 2
| b_\epsilon | \right) \partial_x^2 - \left( b_\kappa + 2 | b_\epsilon | \right)
\partial_y^2 \ ,
\label{eqn:bias_operator}
\end{equation}
where we consider the principal axes as the $xy$-coordinates. This
bias is non-local because it contains partial
derivatives. Furthermore, it is non-isotropic due to the fact that the
derivatives along the principal directions have different bias. Only
when there is no eccentricity bias ($b_\epsilon=0$), we recover
isotropy. The bias in eq.~\eqref{eqn:bias_operator} is a
generalization of the one in \cite{desjacques2008} for peaks with
eccentricity.

\section{Covariance of the multipolar profiles}
\label{sec:covariance}

In this section, the covariance of the multipolar profiles are
calculated. As in the previous section, both the temperature and the
Stokes parameters are expanded in terms of the multipolar profiles. In
the case of peaks with spherical symmetry, only the scalar profile
($m=0$) contributes to the peak local shape. Conversely, if
peaks are selected with eccentricity, then the quadrupolar profile
($m=2$) is also non-zero. In this case, obviously, the multipolar
profiles with $m=0$ and $m=2$ have all the information concerning the
peak shape.

In general, the covariance between the multipolar profiles of the
field $X$ and $Y$ can be written as the sum of two contributions:
\begin{equation}
\langle X_{m^\prime}^* Y_{m} \rangle = \langle X_{m^\prime}^* Y_{m}
\rangle_{\text{intr.}} + \langle X_{m^\prime}^* Y_{m}
\rangle_{\text{peak}} \ ,
\end{equation}
where the intrinsic covariance $\langle X_{m^\prime}^* Y_{m}
\rangle_{\text{intr.}}$ represents the correlations of the multipolar
profiles, independently whether a peak is selected or not. The second
part $\langle X_{m^\prime}^* Y_{m} \rangle_{\text{peak}}$ is a
modification of the intrinsic covariance due to the fact that a peak
is present in the field. In general, the contribution of the peak is a
suppression of the intrinsic covariance caused by the reduction of the
field randomness when the peak variables are constrained.

The intrinsic covariance is given by
\begin{equation}
\langle X_{m^\prime}^*(\theta^\prime) Y_{m}(\theta)
\rangle_{\text{intr.}} = \delta_{mm^\prime}\ \sum_{\ell = m}^\infty
\frac{2\ell+1}{4\pi} \frac{(\ell-m)!}{(\ell+m)!} \ C_\ell^{XY}
\ P_\ell^m(\cos \theta) P_\ell^m(\cos \theta^\prime) \ .
\label{eqn:cov_mm}
\end{equation}
Whilst this covariance is only determined by the angular cross-power
spectrum of the fields and affects to the whole range of $m$-values,
the peak covariance depends on how the peak variables are selected. In
addition to the covariance of the $m=0$ and $m=2$ profiles, the peak
also modifies the covariance of the multipolar profile with $m=1$,
which is associated to the first derivative. The condition of having a
critical point ($\eta=0$) implies that the expected value of the
dipolar profile is zero, and for this reason we have not considered
the $m=1$ profile in the peak shape analysis (see Section
\ref{sec:multipolar}). However, as the covariance of the field is
affected by the constraint on the first derivative, the parameter
$\eta$ must be included in the analysis of this section.

The contribution of the peak to the field covariance is caused by the
particular constraints on the peak degrees of freedom (for instance,
imposing the extremum constraint, the peak height above a given
threshold, or the first derivative equal to zero). These constraints
modify how the peak variables are distributed with respect to the case
without peak selection. In the following, the covariance matrix of
$\nu$, $\kappa$, $\eta$ and $\epsilon$, when peak variables are
unconstrained, is denoted by $S$. Once the peak is selected, the
change in this covariance is parametrized by the matrix $\Delta S$,
which is defined as the difference between the covariance of $\nu$,
$\kappa$, $\eta$ and $\epsilon$, with and without the peak constraints
imposed. The bias of $S$ is defined as the matrix:
\begin{equation}
B_S = B \ (\Delta S ) \ B^\dagger \ ,
\label{eqn:bs}
\end{equation}
where the four-dimensional matrix $B$ is given by the inverse
of $S$, normalizing the rows by the corresponding variances of the peak
variables. That is,
\begin{equation}
B = \left(
\begin{array}{cccc}
\sigma_\nu^{-1} & 0 & 0 & 0 \\
0 & \sigma_\kappa^{-1} & 0 & 0 \\
0 & 0 & \sigma_\eta^{-1} & 0 \\
0 & 0 & 0 & \sigma_\epsilon^{-1} \\
\end{array}
\right) S^{-1} \ .
\end{equation}
The bias of the covariance $B_S$ in eq.~\eqref{eqn:bs} is a linear
transformation of the matrix $\Delta S$. Therefore, if the peak
variables are not constrained (i.e., $\Delta S = 0$), the bias $B_S$ is
also zero. For convenience, the bias matrix $B_S$ is separated in
different blocks taking into account the different spin of the peak
variables:
\begin{equation}
B_S = \left(
\begin{array}{ccc}
B_{00}  & B_{01} & B_{02} \\
B_{10}  & B_{11} & B_{12} \\
B_{20}  & B_{21} & B_{22} \\
\end{array}
\right) \ .
\end{equation}
This matrix is Hermitian by construction, and therefore $B_{ij} =
B_{ji}^*$. The reason of this decomposition is that the peak
variables affect to the different multipolar profiles depending on
their spin. The two-dimensional matrix $B_{00}$ represents the bias of
the covariance of the scalar degrees of freedom ($\nu$ and $\kappa$),
while $B_{11}$ and $B_{22}$ are the biases of the variances of the
first derivative ($\eta$) and the eccentricity ($\epsilon$),
respectively. Likewise, due to the peak selection process, it is
possible to have correlations between different peak variables, which
are described by the off-diagonal terms of $B_S$ (for instance, the
extremum constraint $|\epsilon| \leq |\sqrt{a} |\kappa|$ introduces
correlations between $\kappa$ and $\epsilon$). In the particular case
of peaks where the first derivative is set to zero by definition, the
bias in the covariance of $\eta$ is $B_{11} = -1/\sigma_\eta^2$ and
there is no correlation between $\eta$ and the rest of degrees of
freedom, which leads to $B_{01} = B_{12} = 0$. Finally, the peak
covariance is calculated using the bias matrix $B_S$:
\begin{equation}
\langle X_{m^\prime}^*(\theta^\prime) Y_{m}(\theta)
\rangle_{\text{peak}} =
\mathbf{C}_{m^\prime}^{TX}(\theta^\prime)^\dagger
B_{m^\prime m} \mathbf{C}_{m}^{TY}(\theta) \ ,
\label{eqn:cov_mm_peak}
\end{equation}
where $\mathbf{C}_{m}^{TX}$ for $m=0,2$ are defined in
eq.~\eqref{eqn:c0_c2}. In the particular case of $m=1$, this
quantity is given by the covariant derivative of the correlation
function:
\begin{equation}
\mathbf{C}_1^{TX}(\theta) = \slashed{\partial}^* C^{TX}(\theta) \ .
\end{equation}
In eq.~\eqref{eqn:cov_mm_peak}, it is assumed that the peak is selected
in temperature, but it can be generalized for peaks in any other
field replacing $T$ by that field.

\begin{figure}
\begin{center}
\includegraphics[scale=0.58]{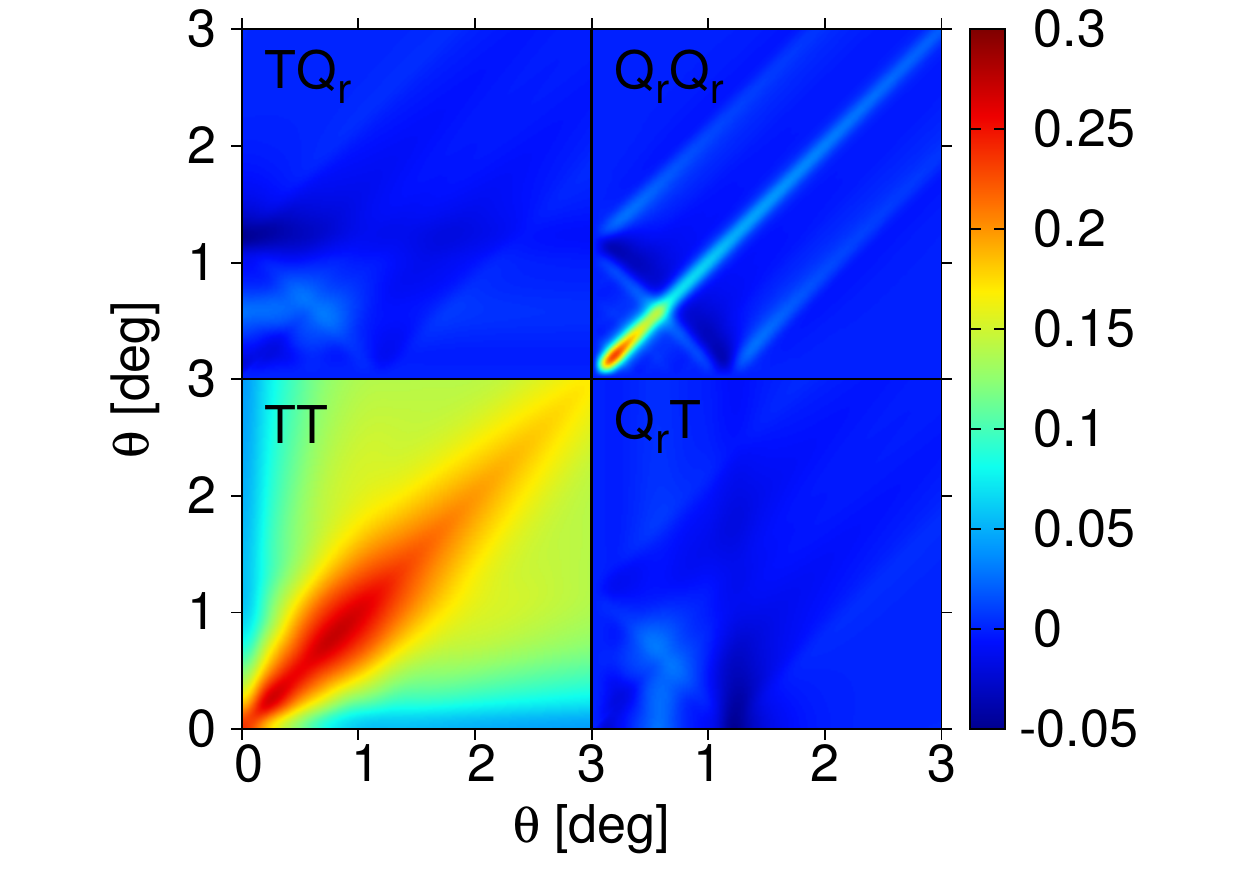}
\includegraphics[scale=0.58]{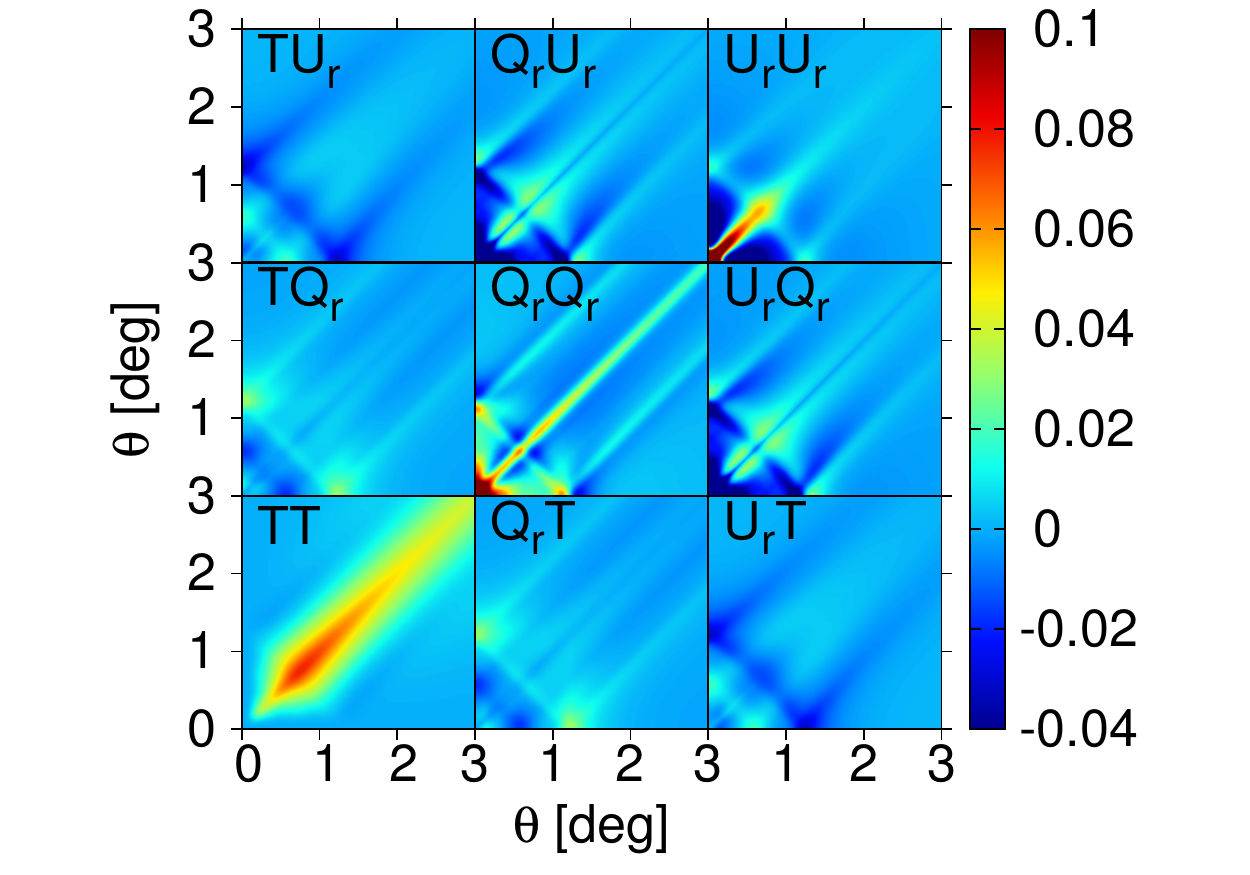}
\end{center}
\caption{Covariance of the $m=0$ (left) and $m=2$ (right) profiles for
  peaks selected in temperature with $\nu > 1$. Each field is
  normalized by the corresponding standard deviation ($\sigma_\nu$ for
  the temperature and $\sigma_P / \sqrt{2}$ for each of the Stokes
  parameters).}
\label{fig:cov_00_22}
\end{figure}

When a peak is present in the field, the covariance is reduced
coherently depending on how the peak variables are constrained. For
instance, if the peak height $\nu$ is fixed to a given value or
selected above a threshold, the field at the centre is constrained,
and therefore it is expected that the variance at $\theta = 0$ is
reduced. In figure~\ref{fig:cov_00_22}, it is represented the
covariance of the $m=0$ and $m=2$ profiles for peaks selected in
temperature with $\nu > 1$. It is possible to see that the effect of
the peak on the covariance mainly affects the $TT$ part, while the
covariances concerning the Stokes parameters are dominated by the
intrinsic term. This fact is produced because the peak covariance of
the Stokes parameters is proportional to the square of the $TE$
correlation, which is subdominant with respect to the intrinsic
fluctuations of the field. This is not the case for temperature, where
the presence of a peak modifies drastically the covariance around the
centre and introduces correlations between different
$\theta$. Additionally, it is possible to consider the covariance
between the monopolar and the quadrupolar profiles. However, the
intrinsic part vanishes and the effect of the peak is small in this
case.

\begin{figure}
\begin{center}
\includegraphics[scale=0.45]{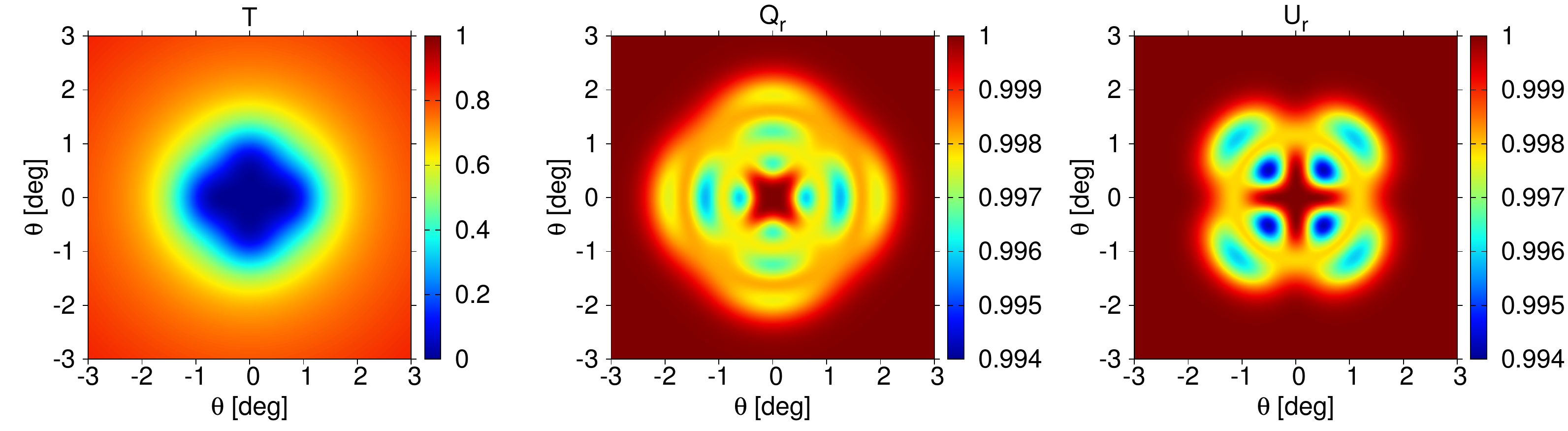}
\end{center}
\caption{Variance of the temperature (left), $Q_r$ (middle) and $U_r$
  (right) fields around a peak selected in $T$ smoothed by a Gaussian
  of FWHM $1^\circ$. The peak height is conditioned to be $\nu=1$, and
  therefore these variances correspond to the patterns in
  figure~\ref{fig:patch}. Each field is normalized by its variance
  corresponding to the case of no peak selection. It is possible to
  see that the peak barely alters the variance of the Stokes
  parameters ($< 1 \%$), whilst the temperature variance is
  drastically decreased in the region of the peak. }
\label{fig:cov_patch}
\end{figure}

If we are interested in analysing the two-dimensional pattern instead
of the individual multipolar profiles, it is necessary to calculate
the covariance of the field $X(\theta,\phi)$. Since all the
information of the field is contained in the multipolar profiles, the
field covariance can be calculated from the covariance of the
multipolar profiles:
\begin{equation}
 \langle X(\theta^\prime,\phi^\prime) X(\theta,\phi) \rangle =
 \sum_{m,m^\prime=-\infty}^\infty \langle
 X_{m^\prime}^*(\theta^\prime) X_m(\theta) \rangle
 \ e^{i(m\phi-m^\prime\phi^\prime)} \ .
\end{equation}
This covariance can also be split into the intrinsic and the peak
contributions. As it is expected, the intrinsic part obtained from
eq.~\eqref{eqn:cov_mm} leads to the field correlation function,
depending on the separation of the two points. On the other hand, the
peak contribution is modelled by the covariance of the multipolar
profiles with $m=0,1,2$ in eq.~\eqref{eqn:cov_mm_peak}. These terms
introduce a inhomogeneous correlation function around the peak, which
can be also anisotropic if the peak has eccentricity. In
figure~\ref{fig:cov_patch}, it is represented the variance of each
point around an oriented peak selected in temperature. In the region
close to the centre, the variance is suppressed with respect to the
intrinsic variance. The quadrupolar pattern present in this figure is
a consequence of the peak eccentricity. Whilst the peak has a strong
effect on the variance of the temperature field, the variances of the
Stokes parameters are modified in less than $1\%$.

\section{Physical interpretation of the peak patterns}
\label{sec:physical}

The azimuthally averaged temperature peak patterns, where the effect
of the eccentricity has been averaged out at zero, are essentially
given by the correlation function between $T$ and the field where the
pattern is imprinted, which can be either $T$ or the polarization
fields. Modifications due to the peak curvature can arise in the
low-peak limit, but this effect is only manifested in the region close
to the centre of the peak and it has not influence in the physical
behaviour of the profiles at large scales. In the high-peak limit, or
for distances greater than the correlation of $\nabla^2 T$, the
physics of peaks is the same as the one causing the shape of the
corresponding temperature cross-correlation functions. For instance,
the ring structure seen in the $Q_r$ profile
(figure~\ref{fig:profile_qr0}) is an effect of the baryon acoustic
oscillations produced at scales smaller than the sound horizon size at
the decoupling epoch, which are also present in the $TQ_r$ (or $TE$)
correlation \cite{komatsu2011,planck162015}.

In this paper, we analyse the effect of the eccentricity in both
temperature and polarization patterns. In the case of the temperature,
the eccentricity of the peak affects to the second order derivatives
at the centre adding a directional dependence. This effect modifies
essentially the small scales since the eccentricity term is
proportional to $\sim \ell^2$. However, the eccentricity is noticeable
at scales up to the sound horizon size. The acoustic oscillations
produced inside of a non-spherical potential propagate the anisotropy
from the centre to the horizon size. In contrast, for scales greater
than the horizon, the physics is dominated by gravity, which is not
sensitive to the local geometry of the potential well, and the
spherical symmetry is therefore recovered. In
figure~\ref{fig:profile_t2}, we can see this effect, where the peak
profile is represented for different azimuthal angle. The quadrupolar
profile $T_2(\theta)$ characterizes the effect of the eccentricity on
the temperature peak as a function of $\theta$. The eccentricity does
not alter the peak height at the centre, and therefore $T_2(\theta)$
vanishes at $\theta=0$. However, this term contributes at scales
within the sound horizon. For larger scales, $T_2(\theta)$ goes to
zero and the peak becomes spherically symmetric.

In addition, the peak orientation in temperature also affects to the
polarization pattern. As described in \cite{komatsu2011}, the
polarization direction characterizes the flow of the photons. Whilst
the polarization direction is radial when the velocity field is
converging, it shows a tangential configuration for a divergent flow
\cite{coulson1994}. In the case of peaks, its shape depends on the
correlation between $T$ and polarization. Therefore, in addition to
the divergence of the photon flow, the sign of the temperature is also
important to describe the polarization pattern. The oscillations in
$Q_r$ represent changes both in the sign of the temperature and in the
velocity field (see \cite{komatsu2011}, for a more detailed
explanation). When the peak has eccentricity, it is possible to
distinguish two different effects on the velocity field which modify
the polarization pattern: a change in the direction of the flow and
azimuthal variations of the modulus of the velocity field. Both
effects modify the local quadrupole moment of the photon distribution,
which causes the CMB polarization. The fact of having a
non-spherically symmetric potential makes the flow to deviate from
being purely radial. This introduces a nonzero $U_{r}$ field, even if
the curl contribution is zero (see figure~\ref{fig:patch}). In the
principal axes directions, the flow is radial as in the spherical
case, and therefore $U_r$ vanishes. However, the deviation from the
radial flow due to the peak deformation reaches its maximum value in
directions at $45^\circ$ with respect to the principal axes. For this
reason, the azimuthal dependence of $U_r$ is a quadrupolar pattern
rotated $45^\circ$ with respect to the orientation axis. The
alternating sign in each quadrupolar lobe indicates that the deviation
angle between the velocity field and the radial direction has
different signs in each quadrant. In addition, the $U_{r}$ pattern
also presents a radial dependence (see
figure~\ref{fig:profile_ur2}). The changes on the sign in the radial
profile is produced by the acoustic oscillations present in the
correlation function of the temperature and polarization fields. In
addition to the flow direction, the modulus of the velocity field is
also affected by the peak eccentricity. In regions where the peak
pattern is compressed with respect to the spherical case, the pressure
of the photons is higher, and on the contrary, the pressure is lower
in the direction of elongation. The pressure of the photon fluid
modifies the velocity field, and hence also the polarization
pattern. The directions of elongation and compression correspond to
the major and minor axes, respectively. This introduces a quadrupolar
pattern aligned with the principal axes of the peak, which can be seen
in both, $Q_r$ and $P$. In some cases, the pressure in the elongation
axis is not enough to reverse the flow, and therefore the change of
sign in $Q_r$ due to the velocity reversion is not present (see
figure~\ref{fig:profile_qr2}).

In order to enhance the elliptical patterns, the peaks represented in
figures~\ref{fig:profile_t2}-\ref{fig:patch_pol} are selected in the
temperature field smoothed with a Gaussian of FWHM $1^\circ$, which
implies that the inner acoustic oscillations in the $Q_r$ profile are
suppressed by the filter (compare with
figure~\ref{fig:profile_qr0}). A calculation of the profiles at high
resolution indicates that any source of power at small scales
different from the baryon acoustic oscillations (e.g. lensing or
noise) produce a smearing of the ring pattern present in the
polarization field due to the fact in this situation the peaks do not
trace properly the potential wells at the last scattering surface.

\section{Peak simulations}
\label{sec:simulations}

In this section, we use the formalism developed in
section~\ref{sec:uncorrelating} to generate constrained simulations
having a peak with given characteristics. For simplicity, we consider
the case in which the peak height $\nu$ is fixed to a given value, but
it is possible to generalize the procedure for random values of
$\nu$. The simulations are generated in the spherical harmonic
space. The first step is to generate the variables $\hat{a}_{\ell m}$
defined in eqs.~\eqref{eqn:alms}, which are given as a linear
combination of the standard spherical harmonic coefficients $a_{\ell
  m}$. This property allows us to consider that the $\hat{a}_{\ell m}$
variables are Gaussian under the assumption that the field where the
peak is selected is also Gaussian. These new variables obtained after
the orthogonalization process are not independent. Their covariance
matrix is given by
\begin{subequations}
\begin{equation}
\langle \hat{a}_{\ell 0} \hat{a}_{\ell^\prime 0} \rangle = 
\delta_{\ell\ell^\prime} + 
\left( \begin{array}{cc}
\nu_{\ell} & \kappa_{\ell}
\end{array} \right)
C^{-1}
\left( \begin{array}{c}
\nu_{\ell^\prime} \\ \kappa_{\ell^\prime}
\end{array} \right) \quad (\ell \neq \ell_{\nu}, \ell_{\kappa}) \ ,
\end{equation}
\begin{equation}
\langle \hat{a}_{\ell 1} \hat{a}_{\ell^\prime 1} \rangle = 
\delta_{\ell\ell^\prime} + 
\frac{\eta_{\ell} \eta_{\ell^\prime}}{\eta_{\ell_{\eta}}^2} \quad
(\ell \neq \ell_{\eta}) \ ,
\end{equation}
\begin{equation}
\langle \hat{a}_{\ell 2} \hat{a}_{\ell^\prime 2} \rangle = 
\delta_{\ell\ell^\prime} + 
\frac{\epsilon_{\ell}
  \epsilon_{\ell^\prime}}{\epsilon_{\ell_{\eta}}^2} \quad (\ell \neq
\ell_{\epsilon}) \ ,
\end{equation}
\begin{equation}
\langle \hat{a}_{\ell m} \hat{a}_{\ell^\prime m} \rangle = 
\delta_{\ell\ell^\prime} \ \ \ (m > 2) \ ,
\end{equation}
\end{subequations}
where $C = PP^t$, being $P$ the pivot matrix defined in
eq.~\eqref{eqn:pivot_matrix}. Using the Cholesky decomposition of the
covariance matrix, it is possible to simulate the $\hat{a}_{\ell m}$
coefficients.

The next step is to simulate the peak variables. Using the probability
in eq.~\eqref{eqn:prob}, we have to put constraints in order to have a
minimum or maximum. In practice, the easiest way to do this is by
using a Montecarlo approach. Conditioning the peak height to $\nu$,
random values of $\kappa$ and $|\epsilon|$ are generated. The
eccentricity $\epsilon$ is generated from the two independent Gaussian
variables which characterize its real and imaginary parts, while the
curvature $\kappa$ is generated as a Gaussian with mean $\rho\nu$ and
variance $1-\rho^2$ (as it can be deduced from
eq.~\eqref{eqn:prob}). If these numbers satisfy the extremum
constraint $|\epsilon| \leq \sqrt{a}|\kappa|$ and $\kappa>0$
($\kappa<0$) for maximum (minimum) selection, these values are
preserved. Otherwise, they are rejected and generated again until
obtaining a pair of values which satisfy the extremum constraint. The
sign of $\kappa$ is chosen to be positive or negative depending
whether we are selecting minima or maxima respectively.

Once the peak variables $\kappa$, $\epsilon$ and the $\hat{a}_{\ell
  m}$ variables are simulated, the standard spherical harmonic
coefficients $a_{\ell m}$ are recovered using
eqs.~\eqref{eqn:alms_inv}. Given a simulation of the temperature, it
is possible to generate the polarization fields $E$ and $B$ correlated
with it. In order to do this coherently, we simulate the spherical
harmonics coefficients $e_{\ell m}$ and $b_{\ell m}$, which correspond
to $E$ and $B$ respectively, following a Gaussian distribution with
mean and variance given in eqs.~\eqref{eqn:elm_blm_cov_peak}. The
influence of the peak in the polarization fields is given by the
correlation between both fields and the temperature.

\begin{figure}
\begin{center}
\includegraphics[scale=0.5]{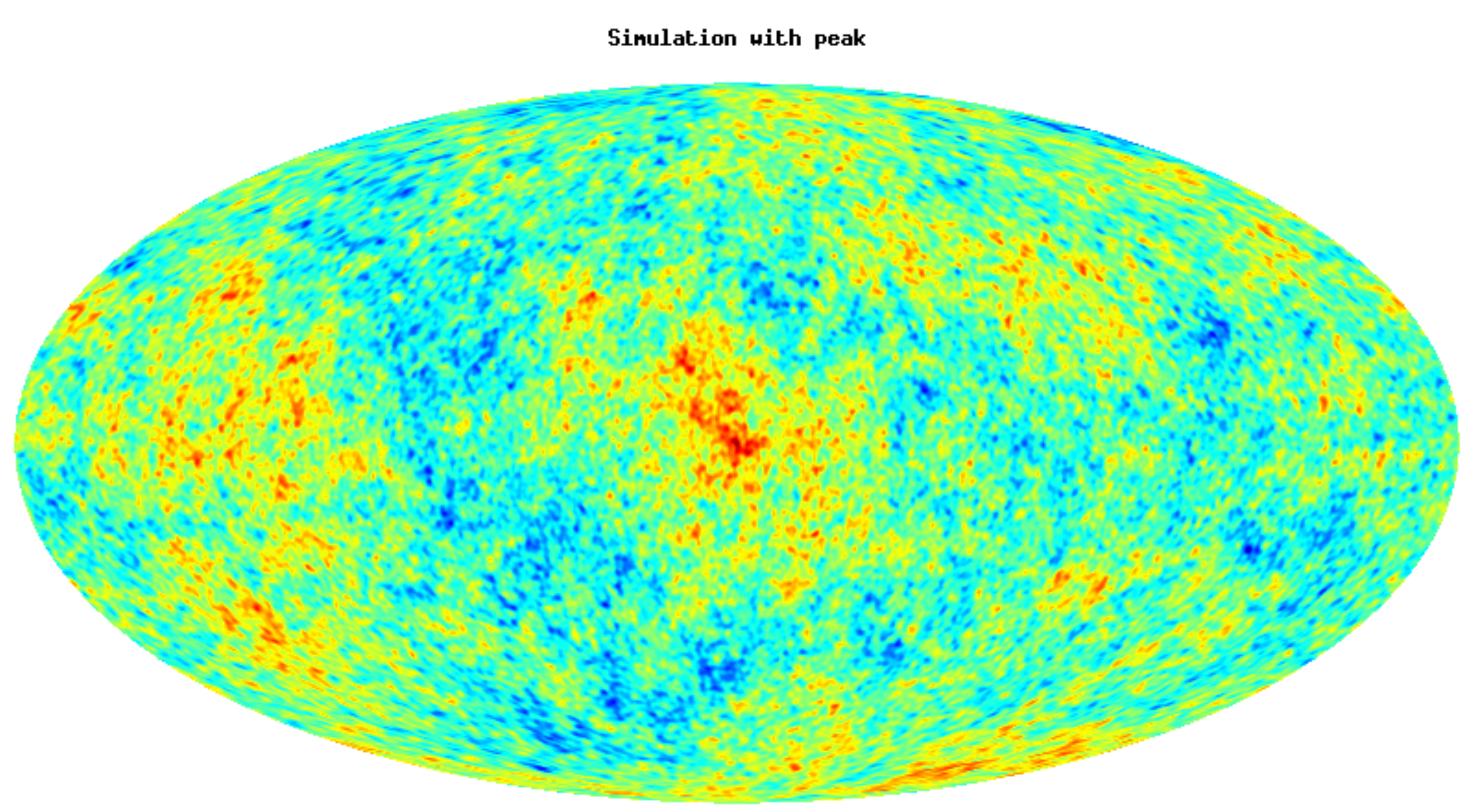}
\includegraphics[scale=0.5]{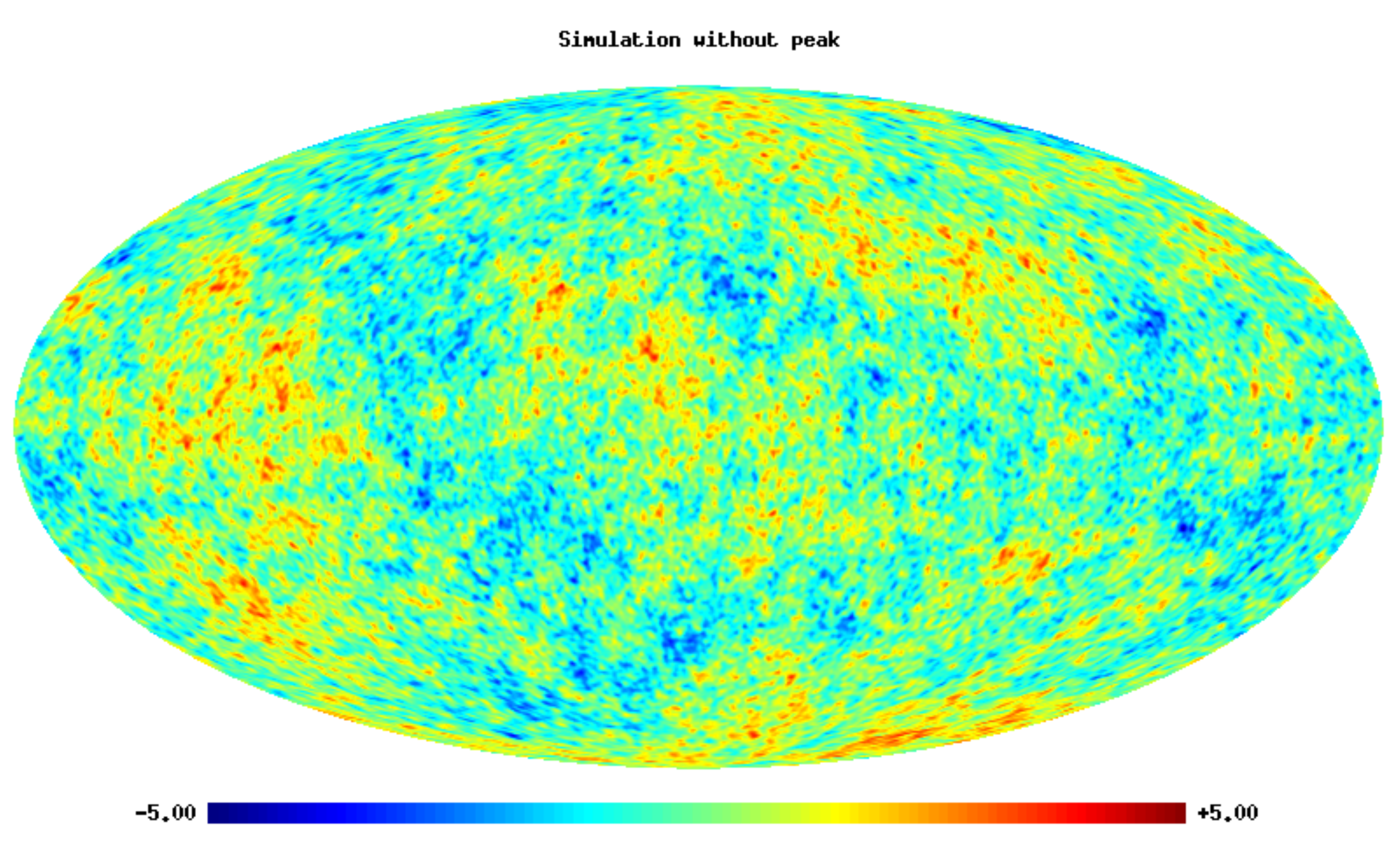}
\end{center}
\caption{Simulations of the CMB temperature field with a peak with
  $\nu=5$ located at the centre of the image (upper map), and without
  a peak (lower map). Both simulations only differ in the peak
  variables (the variables $\hat{a}_{\ell m}$ are the same), and
  therefore it is possible to see similar structures in regions away
  from the peak. One can notice that the presence of the peak affects
  to the area around it, attending to the properties of the
  temperature correlation function. In these maps, the temperature is
  filtered with a Gaussian whose FWHM is $1^\circ$, and the color bar
  indicates the value of the map normalized by the standard
  deviation.}
\label{fig:simulation}
\end{figure}

Notice that although, using this formalism, the peak is located at the
north pole, it can be set at any position on the sphere by performing
the proper rotation. The last step is to construct the maps from the
spherical harmonic coefficients $a_{\ell m}$, $e_{\ell m}$ and
$b_{\ell m}$. A simulation produced following this procedure is given
in figure~\ref{fig:simulation}.

\section{Conclusions}
\label{sec:conclusions}

In this paper, the peak statistics and their shape on the sphere is
presented. The description of the peaks is given by using the suitable
properties of the spherical harmonic space. For this purpose, the peak
degrees of freedom are expressed in terms of the spherical harmonics
coefficients. The peak variables, and the rest of the degrees of
freedom of the field are subject to a decorrelation procedure,
allowing an independent treatment of the peak and the rest of the
fluctuating random field. In this procedure, the assumption of
Gaussianity is essential, since the decorrelation does not guarantee
statistical independence for non-Gaussian fields. The different peak
shapes are obtained taking the expectation value of the random field,
fixing the peak variables to the desired values.

The probability density of the peak variables is also calculated for
the sphere. Some differences with respect to previous calculations are
found \cite{bond1987}, which may be important when the field is
dominated by large-scale peaks. The main difference with respect to
the flat case is that the variances of the mean curvature ($\kappa$)
and the eccentricity ($\epsilon$) are not exactly the same. However,
these variances are not independent since they are related through a
constraint equation (see eq.~\eqref{eqn:var_relation}). In the
small-scale limit, both variances have the same behaviour (they scale
as $\ell^4$) and the flat approximation is recovered. On the contrary,
the variance of the eccentricity is suppressed with respect to the
variance of the curvature for large-scale peaks. Therefore, the
probability density of $\kappa$ and $\epsilon$ is modified for large
peaks on the sphere. However, this effect is only noticeable when the
field is dominated by peaks whose size is $\gtrsim 45^\circ$. Although
these scales are not usually considered, it may be important in the
study of the large-scale anomalies.

The peak shape in $T$, $E$ and $B$ fields for peaks selected in
temperature can be understood as a biased version of the $TT$, $TE$
and $TB$ correlation functions respectively. For high peaks, this bias
is just a constant. However, when the peak height becomes smaller, the
effect of the extremum constraint (minimum or maximum selection) and
the peak eccentricity introduce a non-local bias. It is found that
this bias is anisotropic due to the eccentricity. In the case that
peaks are selected with spherical symmetry, then the non-local
isotropic bias is recovered.

Throughout this work, we consider peaks selected in the temperature
field allowing nonzero eccentricity. The non-spherical symmetry of
peaks introduces a quadrupolar dependence on the azimuthal angle
$\phi$, which modifies their local shape. However, this asymmetry only
affects to scales smaller than the sound horizon size. For larger
scales, the peak shape is only affected by gravity, which is not
sensitive to the local eccentricity at the centre. As it is expected,
although the peaks are selected in the temperature field, the
polarization around the peak location is also affected due to their
correlation. The induced shape on the Stokes parameters, and on the
$E$ and $B$ polarization fields, has been calculated for the general
case of peaks with eccentricity. In the case of the Stokes parameters,
we have used the polar coordinates around the peak, leading to the
parameters $Q_r$ and $U_r$ \cite{kamionkowski1997}. When peaks have
spherical symmetry, and there is no physical effects introducing $TB$
correlation, the induced $U_r$ pattern vanishes. However, this is not
the case when the peak eccentricity is considered. The asymmetrical
photon flow converging or diverging to the potential well introduces a
nonzero $U_r$ contribution. The shape of $U_r$ in this case is a
quadrupole whose axes form an angle of $45^\circ$ with respect to the
peak principal axes. In addition, the Stokes parameter $Q_r$ is also
modified by the peak eccentricity. The differences in pressure and
flow velocity in the directions of elongation and compression of the
ellipse introduce a quadrupolar dependence in $Q_r$, in this case
aligned with the principal axes.

Finally, the peak formalism in the spherical harmonic space developed
in this paper allows to generate Gaussian random simulations with a
given peak at some position on the sphere. The peak can be chosen with
the desired characteristics (peak height, mean curvature and
eccentricity). In particular, the extremum constraint can be imposed
to the peak variables, generating in this case a minimum or
maximum. This mechanism to simulate peaks may be useful for the
analysis of particular peaks present in the data, taking into account
the possible systematics, noise and mask.

In a future work, we will apply the formalism developed in this paper
to CMB data. In particular, we will test the standard cosmological
model looking at the curvature and eccentricity of extrema, and
considering both temperature and polarization.

\appendix

\section{Covariant derivatives on the sphere}
\label{app:cov_der}

A suitable approach to take derivatives on the sphere is by
using the covariant derivatives. The components of a tensor field on
the sphere can be expressed in the standard orthonormal basis of the
tangent plane, $\mathbf{e}_\theta$ and $\mathbf{e}_\phi$. For
convenience, we change this basis to the helicity basis
$\mathbf{e}_{\pm} = \left( \mathbf{e}_\theta \pm i \mathbf{e}_\phi
\right)/\sqrt{2}$. The interest of working in the helicity basis is
that the covariant derivatives can be expressed in terms of the
raising and lowering operators $\slashed{\partial}$ and
$\slashed{\partial}^*$:
\begin{equation}
\nabla_{+} = - \frac{1}{\sqrt{2}} \slashed{\partial} \ , \qquad
\nabla_{-} = - \frac{1}{\sqrt{2}} \slashed{\partial}^* \ .
\end{equation}
Throughout the paper, we use $\slashed{\partial}$ and
$\slashed{\partial}^*$ as the derivative operators instead of the
covariant derivatives, although the difference between both is only a
normalization constant. In order to differentiate any field on the
sphere, it is enough to see how $\slashed{\partial}$ and
$\slashed{\partial}^*$ operate over the spin-weighted spherical
harmonics:
\begin{subequations}
\begin{equation}
\slashed{\partial} \left( {}_{s}Y_{\ell m} \right) =
\sqrt{\ell(\ell+1)-s(s+1)} \ {}_{s+1}Y_{\ell m} \ ,
\end{equation}
\begin{equation}
\slashed{\partial}^* \left( {}_{s}Y_{\ell m} \right) = -
\sqrt{\ell(\ell+1)-s(s-1)} \ {}_{s-1}Y_{\ell m} \ .
\end{equation}
\end{subequations}
For simplicity, we are particularly interested in the value of the
derivatives at the north pole. As the spherical coordinates present
singularities at both poles, we have to take special care when
expressions are evaluated at these points. The problem with the
spherical coordinates is that, while $\theta$ takes the values $0$ or
$\pi$ at the poles, the azimuthal angle $\phi$ is undetermined at
these points. Different values of $\phi$ correspond to different
orientations of the basis vectors $\mathbf{e}_\theta$ and
$\mathbf{e}_\phi$. Therefore, the value of $\phi$ at the poles
characterizes the orientation of the local system of reference. In the
case of scalars, the system of reference is not important due to their
invariant character. However, for higher order tensors, the
orientation modifies their components. In general, if we operate with
$\slashed{\partial}$ and $\slashed{\partial}^*$ over the spherical
harmonics and evaluate them at the north pole ($\theta=0$), it is
obtained that
\begin{equation}
(\slashed{\partial}^*)^a \ (\slashed{\partial})^b \ Y_{\ell m}(0,\phi) =
  (-1)^{b} \sqrt{\frac{2\ell+1}{4\pi}}
  \sqrt{\frac{(\ell+a-b)!}{(\ell-a+b)!}}
  \frac{(\ell+b)!}{(\ell-b)!} \ e^{i(a-b)\phi} \ \delta_{m,a-b}
  \ ,
\label{eqn:sh_der}
\end{equation}
where the $\phi$ dependence has been considered. The spinorial
character of the derivatives causes that their values are complex
numbers. As it is expected, the spin of $(\slashed{\partial}^*)^a
\ (\slashed{\partial})^b \ Y_{\ell m}$ is $a-b$. This fact is
reflected in the exponential factor $e^{i(a-b)\phi}$, which determines
its transformation under azimuthal rotations. The presence of $\phi$
in eq.~\eqref{eqn:sh_der} is nothing more than an indication of the
non-zero spin of the derivatives and the ambiguity of the coordinates
at the north pole. For this reason, we can understand the $\phi$ angle
in this equation as a gauge parameter, caused by the lack of
one-to-one mapping of the spherical coordinates and the sphere. In the
following and throughout the calculations in this paper, we use the
gauge $\phi = 0$ when we evaluate spinorial quantities at the north
pole. This corresponds to a particular orientation of the system of
reference, aligned with the $x$ and $y$ directions. In this case, we
can ignore the factor $e^{i(a-b)\phi}$ in eq.~\eqref{eqn:sh_der}. In
particular, we are interested in some special values of
eq.~\eqref{eqn:sh_der}:
\begin{subequations}
\begin{equation}
Y_{\ell m} (0,0) = \sqrt{\frac{2\ell+1}{4\pi}} \ \delta_{m0}
\end{equation}
\begin{equation}
\slashed{\partial}^* Y_{\ell m} (0,0) = \sqrt{\frac{2\ell+1}{4\pi}}
\sqrt{\frac{(\ell+1)!}{(\ell-1)!}} \ \delta_{m1}
\end{equation}
\begin{equation}
\slashed{\partial} Y_{\ell m} (0,0) = - \sqrt{\frac{2\ell+1}{4\pi}}
\sqrt{\frac{(\ell+1)!}{(\ell-1)!}} \ \delta_{m-1}
\end{equation}
\begin{equation}
\slashed{\partial}^* \slashed{\partial} Y_{\ell m} (0,0) =
- \sqrt{\frac{2\ell+1}{4\pi}} \frac{(\ell+1)!}{(\ell-1)!} \ \delta_{m0}
\end{equation}
\begin{equation}
(\slashed{\partial}^*)^2 Y_{\ell m} (0,0) =
\sqrt{\frac{2\ell+1}{4\pi}} \sqrt{\frac{(\ell+2)!}{(\ell-2)!}}
\ \delta_{m2}
\end{equation}
\begin{equation}
(\slashed{\partial})^2 Y_{\ell m} (0,0) =
\sqrt{\frac{2\ell+1}{4\pi}} \sqrt{\frac{(\ell+2)!}{(\ell-2)!}}
\ \delta_{m-2}
\end{equation}
\end{subequations}

Finally, in order to calculate the Stokes parameters, it is useful to
obtain the expressions for the $2$-spin spherical harmonics, in
particular for $m=0$ and $m=2$:\footnote{The spherical harmonics for
  $m=-2$ are calculated using the property ${}_{\pm s}Y_{\ell -m} =
  (-1)^{m+s} {}_{\mp s}Y_{\ell m}^*$.}
\begin{subequations}
\begin{equation}
{}_{\pm 2}Y_{\ell 0} (\theta,\phi) = Y_{\ell \pm 2} (\theta,\phi)
\ e^{\mp i 2 \phi} = \sqrt{\frac{2\ell+1}{4\pi}}
\sqrt{\frac{(\ell-2)!}{(\ell+2)!}}  \ P_\ell^{2} (\cos \theta) \ ,
\end{equation}
\begin{equation}
{}_{\pm 2}Y_{\ell 2} (\theta,\phi) = 2 \sqrt{\frac{2\ell+1}{4\pi}}
\frac{(\ell-2)!}{(\ell+2)!}  \left( P_\ell^{+} (\cos \theta) \pm
P_\ell^{-} (\cos \theta) \right) \ e^{i2\phi} \ ,
\end{equation}
\end{subequations}
where the functions defined in eqs.~\eqref{eqn:pl+_pl-} were used.

\section{Peak degrees of freedom}
\label{app:peak_dof}

In this appendix, we study the peak degrees of freedom and their
connection to the operators defined in the previous appendix. Peaks
are described by derivatives up to second order. Assuming that the
field is given by its spherical harmonic expansion (see
eq.~\eqref{eqn:spherical_harmonic_expansion}), and that the peak is
located at the north pole, only the $m=0$ spherical harmonic
coefficients contribute to the value of $\nu$
(eq.~\eqref{eqn:nu_kappa}). However, the first derivatives of $T$ at
the north pole are given by the real and imaginary parts of
$\slashed{\partial}^*T$, which is a linear combination of the
spherical harmonics coefficients with $m=1$
(eq.~\eqref{eqn:temp_der}). The second order derivatives are encoded
in the Hessian matrix, which can be written in the following way:
\begin{equation}
\left( \begin{array}{cc}
\partial_x^2T & \partial_x\partial_yT \\
\partial_x\partial_yT & \partial_y^2T
\end{array} \right) = \frac{1}{2}
\left( \begin{array}{cc}
\nabla^2T & 0 \\
0 & \nabla^2T
\end{array} \right) + \frac{1}{2}
\left( \begin{array}{cc}
\mathrm{Re} \ (\slashed{\partial}^*)^2 T & - \mathrm{Im} \ (\slashed{\partial}^*)^2 T \\
- \mathrm{Im} \ (\slashed{\partial}^*)^2 T & - \mathrm{Re} \ (\slashed{\partial}^*)^2 T
\end{array} \right) \ ,
\label{eqn:hessian}
\end{equation}
where $\nabla^2T$ is the Laplacian corresponding to the trace of the
Hessian matrix. It can be written in terms of the operators described
in the previous appendix: $\nabla^2T = \slashed{\partial}^*
\slashed{\partial} T$. The complex number $(\slashed{\partial}^*)^2T$
is the traceless part and it describes the eccentricity of the
peak. The Hessian matrix is separated in this form because the two
parts transform in a different way. The Laplacian is invariant under
rotations around the origin, while the $(\slashed{\partial}^*)^2T$
transforms like a spin-$2$ tensor. The physical meaning of the
Laplacian is the mean curvature of the peak when it is averaged over
all directions. Whilst the modulus of $(\slashed{\partial}^*)^2T$ is
proportional to the square of the eccentricity of the peak, the
orientation angle is encoded in its phase.

Throughout this paper, the peak variables $\nu$, $\kappa$, $\eta$ and
$\epsilon$ are used. They are defined as the quantities $T$,
$-\nabla^2T$, $\slashed{\partial}^*T$ and $(\slashed{\partial}^*)^2T$,
normalized to unit variance. The variances of the peak variables are
\begin{subequations}
\begin{equation}
\sigma_\nu^2 = \langle T^2 \rangle = \sum_{\ell=0}^\infty
\frac{2\ell+1}{4\pi} C_\ell^{TT} \ ,
\end{equation}
\begin{equation}
\sigma_\kappa^2 = \langle (-\nabla^2T)^2 \rangle =
\sum_{\ell=0}^\infty \frac{2\ell+1}{4\pi} \left[
  \frac{(\ell+1)!}{(\ell-1)!} \right]^2 C_\ell^{TT} \ ,
\end{equation}
\begin{equation}
\sigma_\eta^2 = \langle |\slashed{\partial}^* T|^2 \rangle = \sum_{\ell=1}^\infty
\frac{2\ell+1}{4\pi} \frac{(\ell+1)!}{(\ell-1)!} C_\ell^{TT} \ ,
\end{equation}
\begin{equation}
\sigma_\epsilon^2 = \langle |(\slashed{\partial}^*)^2 T|^2 \rangle = \sum_{\ell=2}^\infty
\frac{2\ell+1}{4\pi} \frac{(\ell+2)!}{(\ell-2)!} C_\ell^{TT} \ .
\end{equation}
\end{subequations}
In previous works \cite{bond1987,komatsu2011}, it was
implicitly assumed that $\sigma_\kappa^2$ and $\sigma_\epsilon^2$ are
equal, but we show that they are different if the exact calculation on
the sphere is done. In particular, both variances have a $\ell^4$
behavior at small scales ($\ell \gg 1$) and thus they tend to be
equal. On the contrary, if the field is dominated by large-scale
fluctuations, the variances $\sigma_\kappa^2$ and $\sigma_\epsilon^2$
are different and this has an effect on the peak statistics. In
addition, these two variances are not independent, as they are related
through the following equation:
\begin{equation}
\sigma_\kappa^2 - \sigma_\epsilon^2 = 2 \sigma_\eta^2 \ .
\label{eqn:var_relation}
\end{equation}
In the limit when $\sigma_\kappa^2, \sigma_\epsilon^2 \gg
\sigma_\eta^2$ it is possible to consider that the variances
$\sigma_\kappa^2$ and $\sigma_\epsilon^2$ are equal. The peak
variables $\nu$, $\kappa$, $\eta$ and $\epsilon$ are obtained
normalizing by the respective variance. In this situation, the
multipolar coefficients of the peak variables are
\begin{subequations}
\label{eqn:peak_dof}
\begin{equation}
\nu_\ell = \sqrt{\frac{2\ell+1}{4\pi}
  \frac{C_\ell^{TT}}{\sigma_\nu^2}} \ ,
\label{eqn:nul}
\end{equation}
\begin{equation}
\kappa_\ell = \sqrt{\frac{2\ell+1}{4\pi}
  \frac{C_\ell^{TT}}{\sigma_\kappa^2}}
\frac{(\ell+1)!}{(\ell-1)!} \ ,
\end{equation}
\begin{equation}
\eta_\ell = \sqrt{\frac{2\ell+1}{4\pi}
  \frac{C_\ell^{TT}}{\sigma_\eta^2}}
\sqrt{\frac{(\ell+1)!}{(\ell-1)!}} \ ,
\end{equation}
\begin{equation}
\epsilon_\ell = \sqrt{\frac{2\ell+1}{4\pi}
  \frac{C_\ell^{TT}}{\sigma_\epsilon^2}}
\sqrt{\frac{(\ell+2)!}{(\ell-2)!}}  \ ,
\label{eqn:epsl}
\end{equation}
\end{subequations}
where the factor $C_\ell^{TT}$ has been introduced in order to have
normalized $a_{\ell m}$ coefficients.  It is useful to calculate the covariance
of the peak height, the mean curvature, the first derivative and the
eccentricity. The covariance of $\nu$, $\kappa$, $\eta$ and $\epsilon$
is:\footnote{The covariance of complex variables $\{ x_i \}$ is defined
  as $\langle x_i^* x_j \rangle - \langle x_i^* \rangle \langle x_j
  \rangle$.}
\begin{equation}
S = \left(
\begin{array}{cccc}
1    & \rho & 0 & 0 \\
\rho &    1 & 0 & 0 \\
   0 &    0 & 1 & 0 \\
   0 &    0 & 0 & 1 \\
\end{array}
\right) \ ,
\label{eqn:s}
\end{equation}
where $\rho = \sigma_\eta^2 / \sigma_\nu \sigma_\kappa$ is the
correlation between $\nu$ and $\kappa$. We also consider the
covariance of the scalar degrees of freedom $\nu$ and $\kappa$, which
is a submatrix of $S$:
\begin{equation}
\Sigma = \left(
\begin{array}{cc}
1    & \rho \\
\rho &    1 \\
\end{array}
\right) \ .
\label{eqn:sigma}
\end{equation}

\section{Flat approximation}
\label{app:flat_approx}

In this appendix we see how to calculate the small-angle limit of the
expressions developed throughout this paper. In particular, the
expressions given in \cite{komatsu2011} are recovered. The dictionary
between the sphere and the flat approximation is given by
\begin{subequations}
\begin{equation}
(-1)^m \sqrt{\frac{(\ell-m)!}{(\ell+m)!}}  P_\ell^m(\cos\theta) \quad
  \longrightarrow \quad J_m(\ell\theta) \ ,
\end{equation}
\begin{equation}
\sum_{\ell=0}^\infty \frac{2\ell+1}{4\pi} \quad \longrightarrow \quad
\int \frac{\mathrm{d}\ell}{\ell} \ \frac{\ell(\ell+1)}{2\pi} \ ,
\end{equation}
\begin{equation}
\ell(\ell+1) \quad \longrightarrow \quad \ell^2 \ .
\end{equation}
\end{subequations}
These transformations are valid for large multipoles ($\ell \gg 1$)
and small angles ($\theta \ll 1$) such that $\ell \theta \sim 1$. In
this case, the associated Legendre functions $P_\ell^m(\cos\theta)$
are replaced by the Bessel function $J_m(\ell\theta)$ of order
$m$. This relation can be deduced applying the small-angle limit to
the fundamental equation of the Legendre functions. The sums over
multipoles are replaced by an integral over $\ell$ with the appropriate
volume factor.